\documentclass[iop,apj]{emulateapj}
\usepackage{multirow}
\usepackage{longtable}
\usepackage{ulem}
\usepackage{color}
\usepackage{lipsum}
\usepackage{amsmath}

\newcommand{\eqqref}[1]{Equation (\ref{#1})}
\newcommand{\tabref}[1]{Table~\ref{#1}}
\newcommand{\figref}[1]{Figure~\ref{#1}}
\newcommand{\secref}[1]{Section~\ref{#1}}

\newcommand{\SNeIa}{SNe~Ia}
\newcommand{\SNIa}{SN~Ia}
\newcommand{\C}[1]{\ensuremath{{}^{#1}{\rm C}}}
\newcommand{\Ox}[1]{\ensuremath{{}^{#1}{\rm O}}}
\newcommand{\Ne}[1]{\ensuremath{{}^{#1}{\rm Ne}}}
\newcommand{\Na}[1]{\ensuremath{{}^{#1}{\rm Na}}}

\newcommand{\Ni}[1]{\ensuremath{{}^{#1}{\rm Ni}}}

\newcommand{\Si}[1]{\ensuremath{{}^{#1}{\rm Si}}}
\newcommand{\Fe}[1]{\ensuremath{{}^{#1}{\rm Fe}}}
\newcommand{\code}[1]{\textsc{#1}}
\newcommand{\FLASH}{\code{flash}}
\newcommand{\MESA}{\code{mesa}}

\newcommand{\unitspace}{\ensuremath{\,}}
\newcommand{\usp}{\unitspace}

\newcommand{\unitstyle}[1]{\ensuremath{\mathrm{#1}}}
\newcommand{\power}[2]{\ensuremath{{#1}^{#2}}}

\newcommand{\centi}{\unitstyle{c}}
\newcommand{\kilo}{\unitstyle{k}}

\newcommand{\meter}{\unitstyle{m}}

\newcommand{\second}{\unitstyle{s}}

\newcommand{\Kelvin}{\unitstyle{K}}
\newcommand{\K}{\Kelvin}  

\newcommand{\cm}{\centi\meter}
\newcommand{\gram}{\unitstyle{g}}

\newcommand{\grampercc}{\gram\usp\power{\cm}{-3}} 

\newcommand{\erg}{\unitstyle{ergs}} 

\newcommand{\Msun}{\ensuremath{M_\odot}}

\newcommand{\km}{\kilo\meter}   

\shorttitle{Hybrid Ia Progenitors}

\begin{document}

\title{Type Ia Supernova Explosions from Hybrid Carbon-Oxygen-Neon White Dwarf Progenitors}

\author{
Donald E.\ Willcox\altaffilmark{1},
Dean M.\ Townsley\altaffilmark{2},
Alan C.\ Calder\altaffilmark{3,4},
Pavel A.\ Denissenkov\altaffilmark{5,6},
and Falk Herwig\altaffilmark{5,6}
}

\altaffiltext{1}{
  Department of Physics and Astronomy,
  Stony Brook University, Stony Brook, NY, 11794-3800, USA; \\
  \href{mailto:donald.willcox@stonybrook.edu}{donald.willcox@stonybrook.edu}
}
\altaffiltext{2}{
  Department of Physics and Astronomy,
  The University of Alabama, Tuscaloosa, AL, 35487-0324, USA
}
\altaffiltext{3}{
  Department of Physics and Astronomy,
  Stony Brook University, Stony Brook, NY, 11794-3800, USA
}
\altaffiltext{4}{
  Institute for Advanced Computational Sciences,
  Stony Brook University, Stony Brook, NY, 11794-5250, USA
}
\altaffiltext{5}{
  Department of Physics and Astronomy, University of Victoria, PO Box 1700, STN CSC, 
  Victoria, BC V8W 2Y2, Canada
}
\altaffiltext{6}{
  Joint Institute for Nuclear Astrophysics, Notre Dame, IN 46556, USA
}

\begin{abstract}
Motivated by recent results in stellar evolution that predict the
existence of hybrid white dwarf (WD) stars with a C-O core inside an O-Ne
shell, we simulate thermonuclear (Type Ia) supernovae from these hybrid
progenitors. We use the \FLASH\ code to perform multidimensional simulations
in the deflagration to detonation transition (DDT) explosion paradigm. Our hybrid
progenitor models were produced with the \MESA\ stellar evolution code
and include the effects of the Urca process, and we map the progenitor
model to the \FLASH\ grid. We
performed a suite of DDT simulations over a range of ignition conditions
consistent with the progenitor's thermal and convective structure assuming
multiple ignition points. To compare the results from these hybrid WD
stars to previous results from C-O white dwarfs, we construct a set of
C-O WD models with similar properties and similarly simulate a suite
of explosions.  We find that despite significant
variability within each suite, trends distinguishing the explosions are
apparent in their \Ni{56} yields and the kinetic properties of the ejecta.
We comment on the feasibility of these explosions as the source of 
some classes of observed subluminous events. 
\end{abstract}

\keywords{hydrodynamics --- nuclear reactions, nucleosynthesis, abundances
--- supernovae: general --- white dwarfs}

\section{Introduction}
\label{sec:intro}
Type Ia supernovae (\SNeIa) are bright stellar explosions that 
produce $\sim 0.6~\Msun$ of radioactive \Ni{56}, the decay of
which powers the light curve and leads to a relation between
the peak brightness of an event and the rate of its decline
from maximum~\citep{phillips:absolute}. This relation
enables \SNeIa\ to be used as ``standard
candles'' for cosmological studies, and this use led to 
the discovery that the expansion of the Universe is accelerating 
due to dark energy \citep{riess.filippenko.ea:observational,
perlmutter.aldering.ea:measurements,leibundgut2001}. 

Despite intense study (driven in part by their application as distance
indicators for cosmology), we still have only an incomplete
understanding of the explosion mechanism and fundamental questions,
such as the likely progenitor system(s), persist. It is widely
accepted that \SNeIa\ result from the thermonuclear explosion of a
white dwarf (WD) composed largely of C and O, with this understanding
going back many
decades~\citep{hoylefowler60,arnett.truran.ea:nucleosynthesis}. The
rapid thermonuclear fusion of C and O in a WD releases enough energy
to unbind it, produces the \Ni{56} necessary to power the light curve,
and explains the lack of H observed in the spectra.

There are, however, several possible progenitor systems for such a
configuration. All models involve a binary system and at least one C-O
WD, which follows from the original association of \SNeIa\ with C-O
burning under degenerate conditions~\citep{hoylefowler60}.
Some proposed systems posit a single white dwarf, the single
degenerate (SD) paradigm, and some posit the merger or collision of
two white dwarfs, the double degenerate (DD) paradigm, and within
these are variations.

The ``classic'' model is the Chandrasekhar-mass model in which a white
dwarf gains mass from a companion, a main sequence or red giant star,
or perhaps a He WD, and a thermonuclear runaway occurs just as it
approaches the Chandrasekhar limiting mass
\citep{hoylefowler60,trucam71,whelaniben73,Nomo84}. Alternately, in
the sub-Chandrasekhar-mass scenario, explosive burning in the accreted
layer triggers a detonation at the surface or in the core of a
lower-mass WD \citep{nomoto80,woosleyweavertaam80,simetal2010}.

The other main class of models is the DD progenitor,
\citep{webbink84,ibentutukov84}, in which two WDs inspiral and merge.
This scenario includes inspiraling pairs, collisions, violent mergers,
and also the ``core-degenerate'' model where the merger takes place in
a common envelope
\citep{raskinetal2009,pakmoretal2011,kashi:2011,pakmoretal2012a,Shenetal12}
Also
see~\citet{hillebrandt.niemeyer:type,howell2011,hillebrandtetal2013,calderetal2013}
for additional discussion.

The observational evidence of one progenitor vs.\ another is
conflicting. \SNeIa\ show a wide range of luminosities and also the
possibility that there are two classes of
progenitor~\citep{MannucciEtAl06,howelletal+09,howell2011}.
Observational and population
synthesis~\citep{Belczynski2005New-Constraints,ruiteretal2011}
arguments suggest that there simply may not be enough
Chandrasekhar-mass progenitors to explain the observed \SNIa\ rate.
There is, however, disagreement over the significance of these
observations~\citep{hachisu:2008} and the suggestion has been made
that instead we do not fully understand the pre-supernova evolution of
the different progenitor systems~\citep{DiStefano:2010}. Certainly
there is disagreement in the interpretation of observations that stems
from uncertainty in the accretion phase of SD
evolution~\citep{hachisu:2010}. Additionally, the oft-cited claim that
the WD in the SD channel would lose mass via nova explosions, thereby
preventing it from reaching the Chandrasekhar mass, is
questioned~\citep{Zorotovic:2011}. Prior work on novae and rapidly
accreting WDs strongly suggests that novae will not be able to
grow~\citep{denissenkovetal2013b,denissenkovetal2014}, especially not
from WD masses of $0.83~\Msun$, suggested to be the mean mass of WDs in
cataclysmic variables in~\citet{Zorotovic:2011}. However, if it is
possible to get a \SNeIa\ out of the SD scenario, then ``hybrid''
C-O-Ne WDs (WDs with a C-O core in an O-Ne
shell)~\citep{denissenkovetal2015} may play a key role (cf.
\secref{section:hybrid_progenitor_models}). These hybrid WDs would
provide ignitable Ia progenitors that are already very close to the
Chandrasekhar limiting mass, and are therefore perhaps the most likely
to produce a \SNeIa.

While there is uncertainty, some contemporary observations do strongly
support the SD progenitor. Events like PTF11kx and others show distinct 
circumstellar shells of material that can be best explained in the SD
context~\citep{dildayetal2012,silverman:2013}. The SNR 3C 397 is heralded
as a case where only an explosion from a Chandrasekhar-mass progenitor can 
produce the nuclei seen in the remnant, due to the need for electron captures 
at high density~\citep{3c397}.  
The recent observation of a UV pulse~\citep{cao:2015} in the
early evolution of an \SNIa\ also supports the SD model. 
Observations of remnants also offer support 
for Chandrasekhar-mass explosions, including wind blown shells in 
RCW86~\citep{williams:2011} and shocked circumstellar material/bubble in the Kepler
remnant~\citep{chiotellis:2012,burkey:2013}.  Altogether, there is
substantial evidence that suggests that the SD channel plays a role in
at least some of the observed \SNeIa~\citep{baron2014}.

The sub-Chandrasekhar-mass model does not have the population synthesis 
arguments
working against it and we know low-mass WDs in binary systems exist. Systems
that are believed will evolve to an explosion have been
observed~\citep{kilic:2014}, potential events have been
identified~\citep{geier:2013,inserra:2015}, and the Type Iax sub-class
of SN Ia~\citep{foleyetal2013,wang:2013} have been suggested as being
sub-Chandrasekhar-mass events themselves.  

Observational evidence also supports the DD progenitor system, and
the scenario is  increasingly seen as the likely progenitor of some
events. SN 2011fe has been intensely observed and does not show features
in its spectra that would be expected if there were a normal stellar
companion~\citep{graham:sn2011fe}, suggesting a DD system. Super-Chandrasekhar
mass explosions like SN~2007if~\citep{scalzo:2010,Yuan:2010} and
SNLS~03D3bb~\citep{howell+06} also suggest mergers. There are
also many population synthesis arguments in favor of mergers as well 
\citep[see][for a review]{maoz:2014}.

\subsection{The Chandrasekhar-mass Single Degenerate Scenario}

In the Chandrasekhar-mass scenario, the central temperature and 
density of the WD increase as it accretes mass from a binary companion 
and approaches the limiting Chandrasekhar mass. As the mass
approaches the limit, central conditions become hot enough for 
carbon fusion to begin (via the \C{12}--\C{12} reaction), driving 
the development of convection throughout the interior of the WD
\citep{Baraffe2004Stability-of-Su,WoosWunsKuhl04,wunschwoosley2004,Kuhletal06,nonakaetal2012}.
As the central temperature reaches $\sim 7\times 10^8~\mathrm{K}$, the
fuel in a convective plume burns to completion before it can cool via
expansion \citep{Nomo84,WoosWunsKuhl04}, and a flame is born. 

The nature of this burning, be it a supersonic detonation or subsonic 
deflagration, largely determines the outcome of the explosion. 
It has been known for some time that a purely supersonic burning front 
cannot
explain observations because the supersonic front very rapidly incinerates 
the star without it having time to react and 
expand~\citep{arnett.truran.ea:nucleosynthesis}. The lack of expansion 
allows most of the star to burn at high densities,
which produces an excessive $^{56}$Ni yield and does not match the
stratified composition of observed remnants~\citep{mazzalietal2008}.  
Instead, a subsonic deflagration must ignite, which allows the
outer layers of the star to expand ahead of the burning front.
In this case, the density of the expanding material decreases, 
which leads to incomplete burning of more mass and thus
increased production of intermediate mass elements. This
deflagration must accelerate via instabilities and turbulent
interaction, a topic that has been explored extensively in the 
past~\citep{khokhlov1993,bychovliberman1995,
SNrt,Khok95,NiemHill95,khoketal1997,ZingDurs07,
cholazarianvishniac2003,
roepkehn2003,roepkehn2004,
Zingale2005Three-dimension,Schmetal06a, Schmetal06b,Aspdetal08,
Woosetal09,csetal2009,hicksrosner2013,c-ssr2013,
jacketal2014,poludnenko2015,hicks2015}.

A deflagration alone will not produce a event of normal brightness and
expansion velocity~\citep{roepkeetal07}. Instead, the initial
deflagration must transition to a detonation after the star has
expanded some in order to produce abundances and a stratified ejecta
in keeping with
observations~\citep{Khokhlov1991Delayed-detonat,hoflich.khokhlov.ea:delayed}.
The physics of this ``deflagration-to-detonation transition'' (DDT)
are not completely understood, but there has been considerable study
based on mechanisms involving flame fronts in highly turbulent
conditions~\citep{1986SvAL, woosley90, Khokhlov1991Delayed-detonat,
  hokowh95, HoefKhok96, khoketal1997, NiemWoos97, hwt98, Niem99,
  GameKhokOran05,roepke07, poletal2011,c-ssr2013,poludnenko2015}.
These models generally reproduce the observations under certain
assumptions about the ignition~\citep{townetal2009}, but research has
shown that the results are very sensitive to the details of the
ignition~\citep{PlewCaldLamb04,GameKhokOran05,garciasenz:2005,
  roepkeetal07,Jordan2008Three-Dimension}.
In our simulations, we
initialize a detonation once the deflagration front reaches a
characteristic DDT fuel density, which controls the degree of
expansion the star undergoes during the deflagration stage. The
implementation details are descibed further in
\secref{subsec:ddt_details}.

\subsection{Systematic Effects}

Contemporary observational campaigns typically investigate how
the brightness and rates of supernovae correlate to 
properties of the host galaxy such as mass and star formation
rate~\citep[c.f.][]{graurmaoz2013,graurbiancomodjaz2015}. Of
particular interest is the delay time distribution (DTD), the
supernovae rate as a function of time elapsed from 
early, rapid
star formation in the host galaxy, and how it may be used
to constrain progenitor models~\citep{hachisu:2008,
conleyetal2011,howell2011,grauretal2011,biancoetal2011,maozmannuccibrandt2012}.
See also the review by~\citet{maozmannucci2012}.
Very recent results indicate evolution of the UV spectrum with 
redshift, providing evidence for systematic effects with
cosmological time \citep{milneandfoley2015}.

Motivated by this interest in correlations between properties of the
host galaxy and the brightness and rate of events, earlier incarnations of  
our group performed suites of simulations in the DDT scenario
with a modified version the \FLASH\ code (described below) to explore
systematic effects on the brightness of an event measured by the yield of
$^{56}$Ni~\citep{Krueger2010On-Variations-o,jacketal2010,kruegetal12}. The
study we present here explores how explosions following from a new class
of ``hybrid" 
progenitors~\citep{denissenkovetal2013,chenetal2014,denissenkovetal2015} 
compares to these previous results.

\section{Hybrid Progenitor Models}\label{section:hybrid_progenitor_models}
Rumors that the structure and evolution of stars is a
  solved problem~\citep{HansenEtAl04} are greatly exaggerated. Recent
  developments obtained with the modern software instrument
  \MESA~\citep{mesa1,mesa2,mesa3} indicate that convective boundary
  mixing (CBM) in the cores of super asymptotic giant branch stars
  (super-AGB) plays a more critical role than previously thought.
  There are several examples in which the use of CBM improves
  agreement between models and observations, including that
  of~\citet{denissenkovetal2013}, which studied WD interior shell
  convection, and~\citet{herwig2005, werner.herwig.2006}, which treated
  He-shell burning in AGB stars.
  \citet{denissenkovetal2013,chenetal2014} found that in some
  super-AGB stars, CBM halts the progression of carbon burning into
  the stellar core, leaving an unburnt C-rich core as large as
  $0.2~\Msun$ surrounded by an O-Ne-rich intershell region extending
  out to the accretion layer at the end of hydrostatic carbon burning. This effect of C-flame quenching via CBM is also confirmed by the extensive parameter study on C-burning in super-AGB stars in~\citet{farmeretal2015}. 

This is the situation in~\citet{denissenkovetal2015}, which explored
the stellar evolution of a super-AGB star with initial mass of
$6.9~\Msun$ and obtained the hybrid white dwarf that is the focus of
the present work. After hydrostatic carbon burning has ceased, the
white dwarf accretes carbon-rich material at its surface, leading to
the rise of temperature near its center. This results in carbon
burning in the upper layer of the small carbon-rich core, which,
together with the thermal effects of the $\Ne{23}/\Na{23}$ Urca process,
provides off-center heating~\citep{denissenkovetal2015} that drives
convection throughout the entire white dwarf except the carbon-rich
core.

\begin{figure}[t]
	\includegraphics[width=\linewidth]{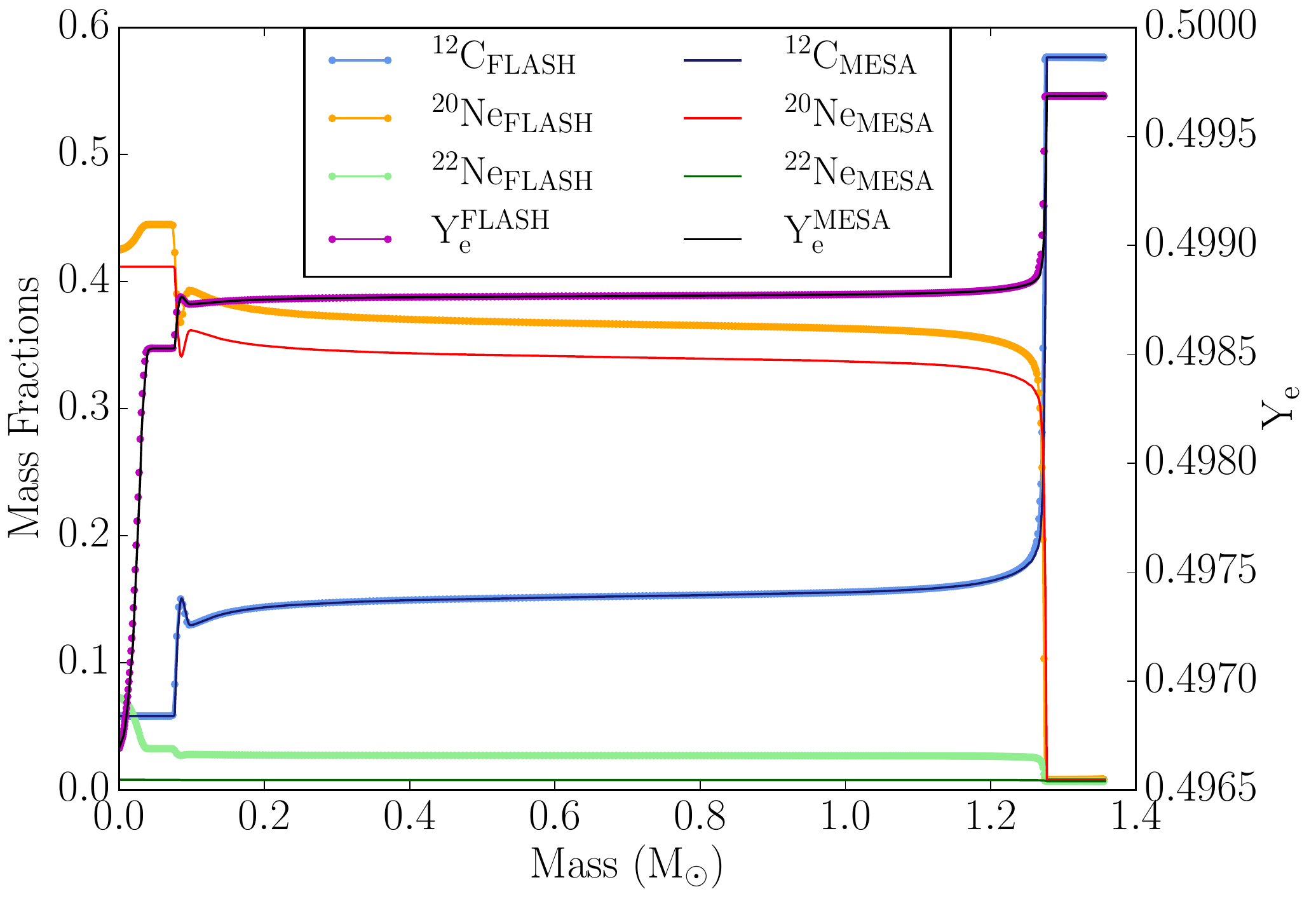}
	\caption{\label{fig:progenitor_abundances} Abundance profile
          of \MESA\ progenitor (\MESA) and its reconstruction on a
          uniform grid at $4$~km spatial resolution with the hydrostatic equilibrium condition of
          \eqqref{eq:hse_pressure} enforced (\FLASH). The reduced set
          of nuclides are shown, where $\mathrm{X_{\Ox{16}} = 1 -
            X_{\C{12}} - X_{\Ne{20}} - X_{\Ne{22}}}$ for the
          abundances labeled \FLASH. Solid lines and circles denote
          the abundances used in \FLASH\ whereas plain solid lines
          denote the abundances used in \MESA.}
\end{figure}

\begin{figure}[t]
	\includegraphics[width=0.9\linewidth]{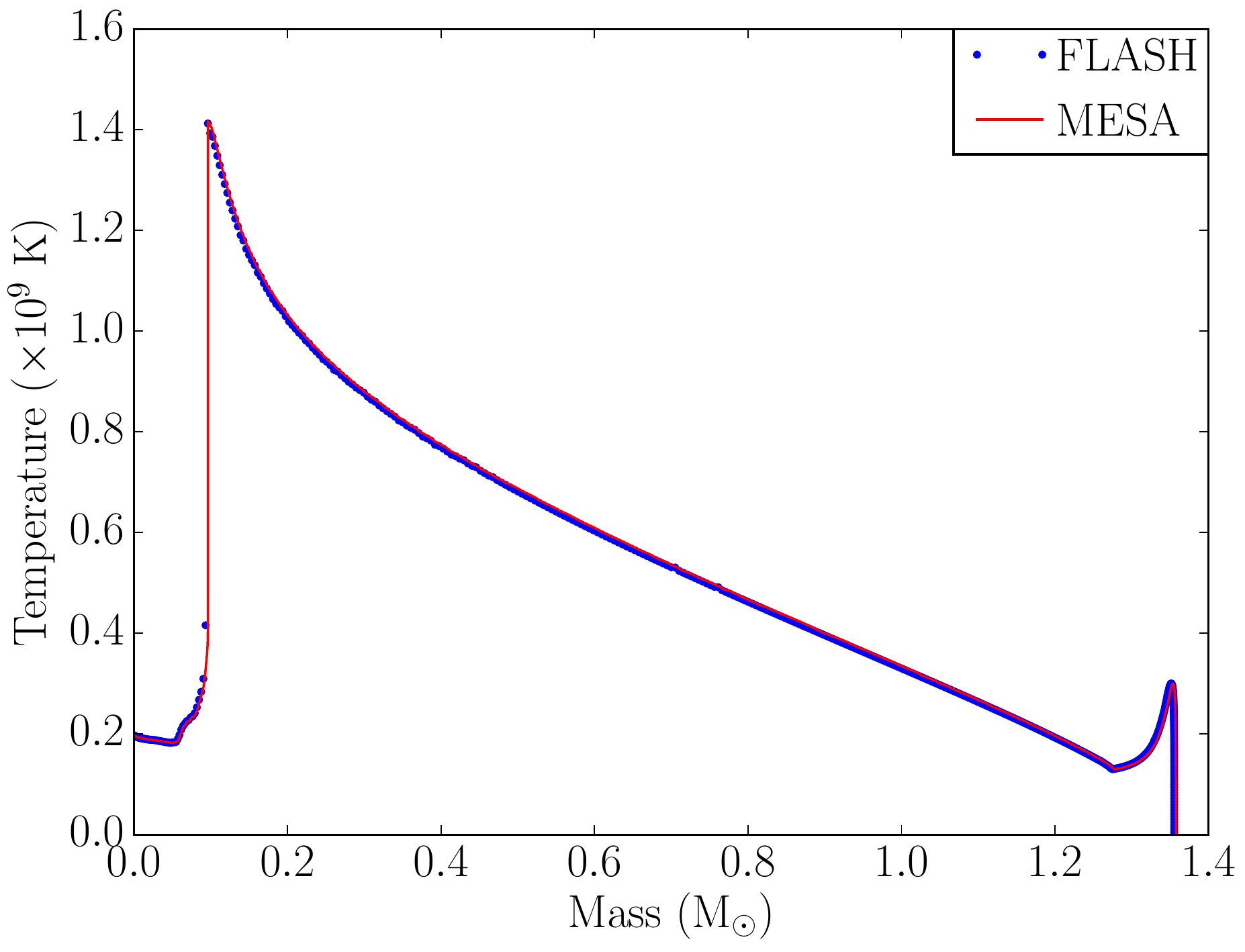}
	\caption{\label{fig:progenitor_temperatures} Temperature
          profile of \MESA\ progenitor (\MESA) shown with a solid red
          line and its reconstruction on a uniform grid at $4$~km spatial resolution (\FLASH) with
          grid zones shown in blue circles.}
\end{figure}

Convection subsequently mixes the carbon-poor material in the
O-Ne intershell region with carbon-rich material on the accreted layer
and also partially mixes carbon-rich material from the core with the
carbon-poor material in the O-Ne intershell. This proceeds along with
accretion and carbon burning, until the latter yields peak
temperatures near $10^9$~\K, around which the local heating time is
shorter than the eddy turnover time so as to ignite thermonuclear
runaway~\citep{wunschwoosley2004}. At this point, the carbon-rich core
has been significantly depleted of carbon and consists mostly of
\Ox{16} and \Ne{20}, while the O-Ne intershell region has been
enriched to a \C{12} abundance of $\approx 0.14$ due to convective
mixing. This scenario, immediately preceding the \SNIa-like explosion,
is shown in \figref{fig:progenitor_abundances} and
\figref{fig:progenitor_temperatures}.  The process of mapping this
\MESA\ progenitor into hydrostatic equilibrium in \FLASH\ is described
in~\secref{subsec:mesa_flash_mapping}.  

This hybrid WD has the interesting property that its mass before the
onset of accretion is $1.06~\Msun$, naturally closer to the
Chandrasekhar limit than a traditional C-O WD. This means that such
hybrid WD progenitors would require less mass accretion to approach
the Chandrasekhar limit, which helps to resolve one of the
difficulties with the SD progenitor
system~\citep{denissenkovetal2015,chenetal2014,kromeretal2015}. The
mass of this hybrid WD following accretion is $1.36~\Msun$.

We note that these models include the influence of the URCA process on the 
convective phase of the pre-explosion progenitor. Our progenitor profiles are
taken directly from MESA models presented in ~\citet{denissenkovetal2015}. 
These include contributions to the energy from thermal energy produced
by the URCA process, but the underlying mixing length theory was not 
modified correspondingly. Thus the effect on the convection is only
due to the energy loss/generation rate. Our progenitor profiles are
shown in Figures \ref{fig:progenitor_abundances} and \ref{fig:progenitor_temperatures}
and correspond to the models in Figure 9a of~\citet{denissenkovetal2015}.
In this regard, this progenitor differs from the carbon-core
models of~\citet{kromeretal2015} in that it includes the pre-explosion convective
burning phase that spreads the carbon enrichment throughout the star before
ignition of the flame front.
While the progenitor we study consists mostly of \Ox{16}
  and \Ne{20}, having the average composition of (\C{12} = 0.17,
  \Ox{16} = 0.42, \Ne{20} = 0.32), it differs from the O-Ne white
  dwarf of~\citet{marquardtetal2015} by having a much higher abundance
  of \C{12} due to accumulation and mixing of \C{12} material during
  the accretion phase, as described above. As discussed in
  \secref{subsec:one_burning_mods} and
  \secref{subsec:hybrid_ignition_parameters}, given the temperature
  profile of \figref{fig:progenitor_temperatures}, this available
  \C{12} is sufficient to drive both a subsonic deflagration and
  subsequently a supersonic detonation front as in previous work that
  applied the same DDT approach to C-O white dwarf
  progenitors~\citep{kruegetal12}.

\section{Methodology}
A few significant new developments in our computational methods were
necessary to simulate the explosion of the hybrid C-O-Ne WD. In
\secref{subsec:one_burning_mods}, we obtain the steady-state
detonation structure for the hybrid C-O-Ne fuel and compare its
detonation characteristics with those of C-O fuel to analyze the
suitability of our combustion model in \FLASH. Then in
\secref{subsec:mesa_flash_mapping} we map the hybrid C-O-Ne WD into a
uniform spatial grid to initialize \FLASH\ while taking care to
preserve hydrostatic equilibrium. We comment in
\secref{subsec:flash_flame} on our \FLASH\ combustion model and in
\secref{subsec:ddt_details} on the deflagration to detonation
transition (DDT) scheme. Finally, we describe the simulation geometry
and the adaptive mesh refinement used in
\secref{subsec:flash_geometry}.

\subsection{Modifications for C-O-Ne Burning} 
\label{subsec:one_burning_mods}

The combustion model in \FLASH\ that we use for Type Ia supernovae
simulations \citep{Caldetal07, townsley.calder.ea:flame,
  SeitTownetal09, townetal2009, jacketal2014, townetal15} separates
the burning into four states: unburned fuel, C-fusion ash, a
silicon-group-dominated nuclear statistical quasi-equilibrium (NSQE)
state, and a full nuclear statistical equilibrium (NSE) state
dominated by iron-group elements (IGEs). The progress of combustion
from one of these states to the next is tracked by three scalar
progress variables whose dynamics is calibrated to reproduce the
timescales of reactions that convert material among these
states. Previous work has focused on fuel mixtures composed
principally of $^{12}$C, $^{16}$O, and $^{22}$Ne.  Simulation of the
hybrid models required extension of this burning model to account for
the presence of \Ne{20} as a large abundance in the fuel. Here we
describe both how $^{20}$Ne is processed during combustion, and the
modifications made to the burning model that accommodate it.

The burning stages above are determined by the hierarchy of timescales
for the consecutive consumption of C and O via fusion and Si via
photo-disintegration and alpha capture. Investigation of the inclusion
of \Ne{20} focused on whether an additional Ne-consumption stage would
be required, and, if not, what stage should include Ne consumption. In
order to characterize the physical burning sequence that we want to
model, we performed a series of simulations of detonations propagating
through WD material with the TORCH nuclear reaction network
software~\citep{timmes1999,torch}.\footnote{TORCH is available from
  \url{http://cococubed.asu.edu}, and the modified version used for
  this study is available from
  \url{http://astronomy.ua.edu/townsley/code}.} TORCH is a general
reaction network package capable of solving networks with up to
thousands of nuclides. A mode is implemented that computes the
one-dimensional spatial thermodynamic and composition structure of a
steady-state planar detonation using the the Zel'dovich, von Neumann,
and D\"{o}ring (ZND) model \citep{FickDavi79,townetal15}. We use a
reaction network composed of 225 nuclides consisting of the 200
nuclides in \citet{Woosley1995The-Evolution-a} in addition to the 25
neutron-rich nuclides added by \citet{Caldetal07} to improve coverage
of electron capture processes in the Fe group.

\begin{figure}[t]
  \includegraphics[width=\linewidth]{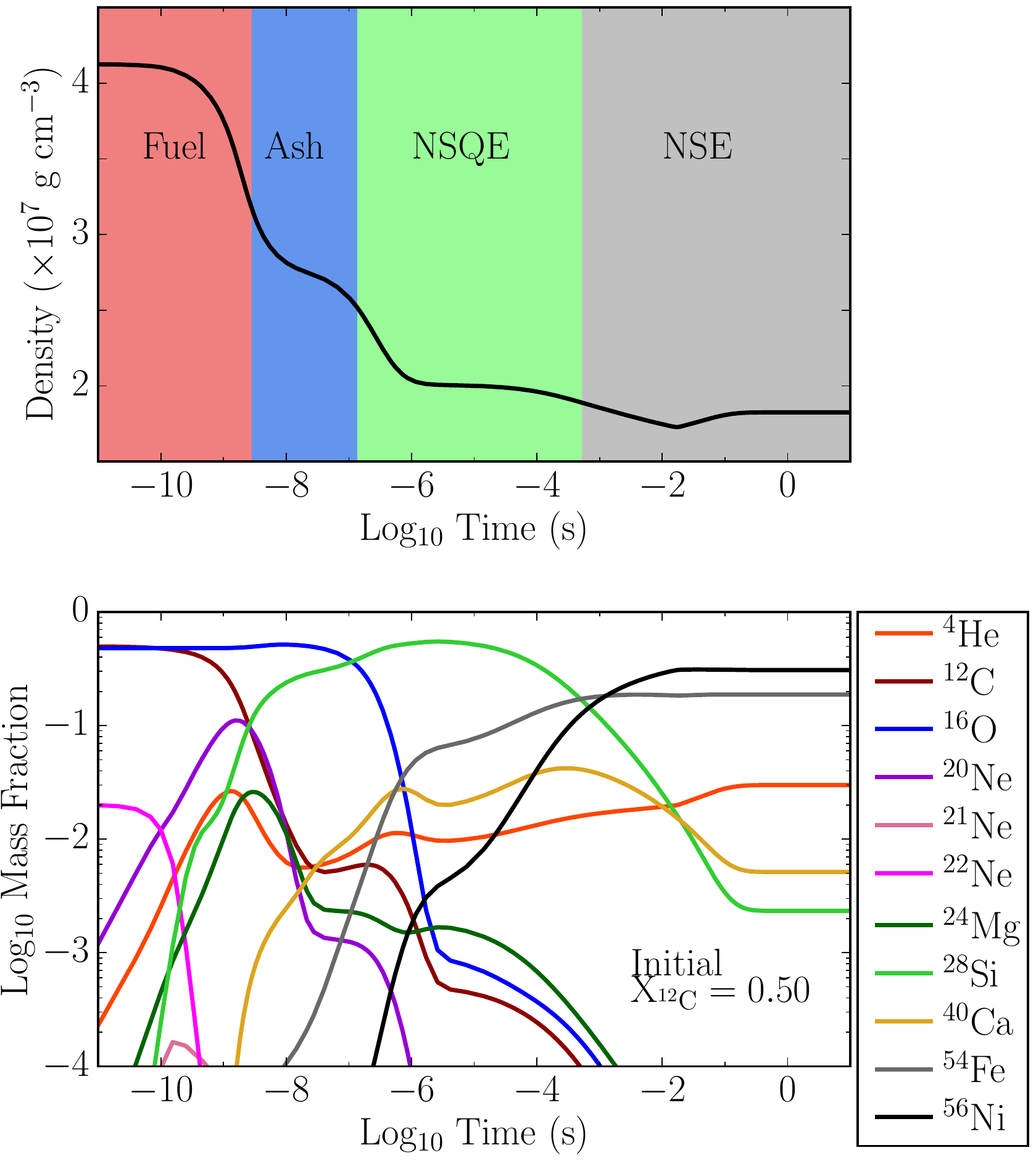}
	\caption{\label{fig:co_znd_stages} Density (upper) and
          abundance (lower) structure in the post-shock flow of a ZND
          detonation calculation of C-O WD material. The upper panel is shaded to show
          the fuel (red), ash (blue), NSQE (green), and NSE (gray) burning stages, demarcated by the \C{12}-\Si{28}, \Ox{16}-\Si{28}, and
          \Si{28}-\Fe{54} crossing times. The initial composition is
          $X_{^{16}\rm O} = 0.48$, $X_{^{22}\rm Ne} = 0.02$,
          $X_{^{12}\rm C} = 0.50$, and $X_{^{20}\rm Ne} = 0.00$, and
          the pre-shock density is $10^7$~g~cm$^{-3}$.}
\end{figure}

For the multi-species fuel and ash relevant to typical white dwarf material, the ZND detonation exhibits the stages that motivate the combustion model.
Figure \ref{fig:co_znd_stages} shows these stages as they appear in a
ZND detonation calculation in fuel with the fractional composition
(\C{12} = 0.50, \Ox{16} = 0.48, \Ne{22} = 0.02) corresponding to
the composition found in the interior of a C-O WD, and a fuel density of $10^7$~g~cm$^{-3}$.
The evolution of the mass fractions in time following the passage of the shock through 
the zone of material is plotted below the density structure in Figure \ref{fig:co_znd_stages}.
The time range of the four states representing the burning stages are indicated by colors in the upper panel of Figure \ref{fig:co_znd_stages}, with the consecutive states separated by the
\C{12}-\Si{28}, \Ox{16}-\Si{28}, and \Si{28}-\Fe{54} crossing times.
It can be seen that the times of the density plateaus are directly comparable to the times at which the primary energy release transitions from one fuel source to another, 
as, e.g., when the \C{12} fraction has fallen
to $\approx1\%$ of its initial value just before $10^{-8}$~s.

In our combustion models for C-O progenitors, consumption of the
initial fuel is modeled as a two-step process. The two stages
represent the consumption first of \C{12}, then of \Ox{16}, mimicking
the sequence seen in the detonation structure shown in
Figure~\ref{fig:co_znd_stages} for ``Fuel'' and ``Ash'' stages. At the
end of this second stage the material is in a \Si{28}-dominated NSQE
state~\citep{Caldetal07,townsley.calder.ea:flame}. To determine how
the burning stages change with the inclusion of \Ne{20}, as is the
case in the hybrid C-O-Ne progenitor, we perform ZND calculations with
an admixture of \Ne{20} ranging from $0.01$ to $0.45$, at the expense
of \C{12} content. For each composition we find the minimally
overdriven solution, as was done in the C-O case. Out to the minimum
in density, this solution is the same as the eigenvalue ZND solution,
which corresponds to a self-supported detonation \citep{FickDavi79,townetal15}.
This computation gives the resolved detonation structure in C-O-Ne Hybrid WD matter.

\begin{figure}
	\includegraphics[width=\linewidth]{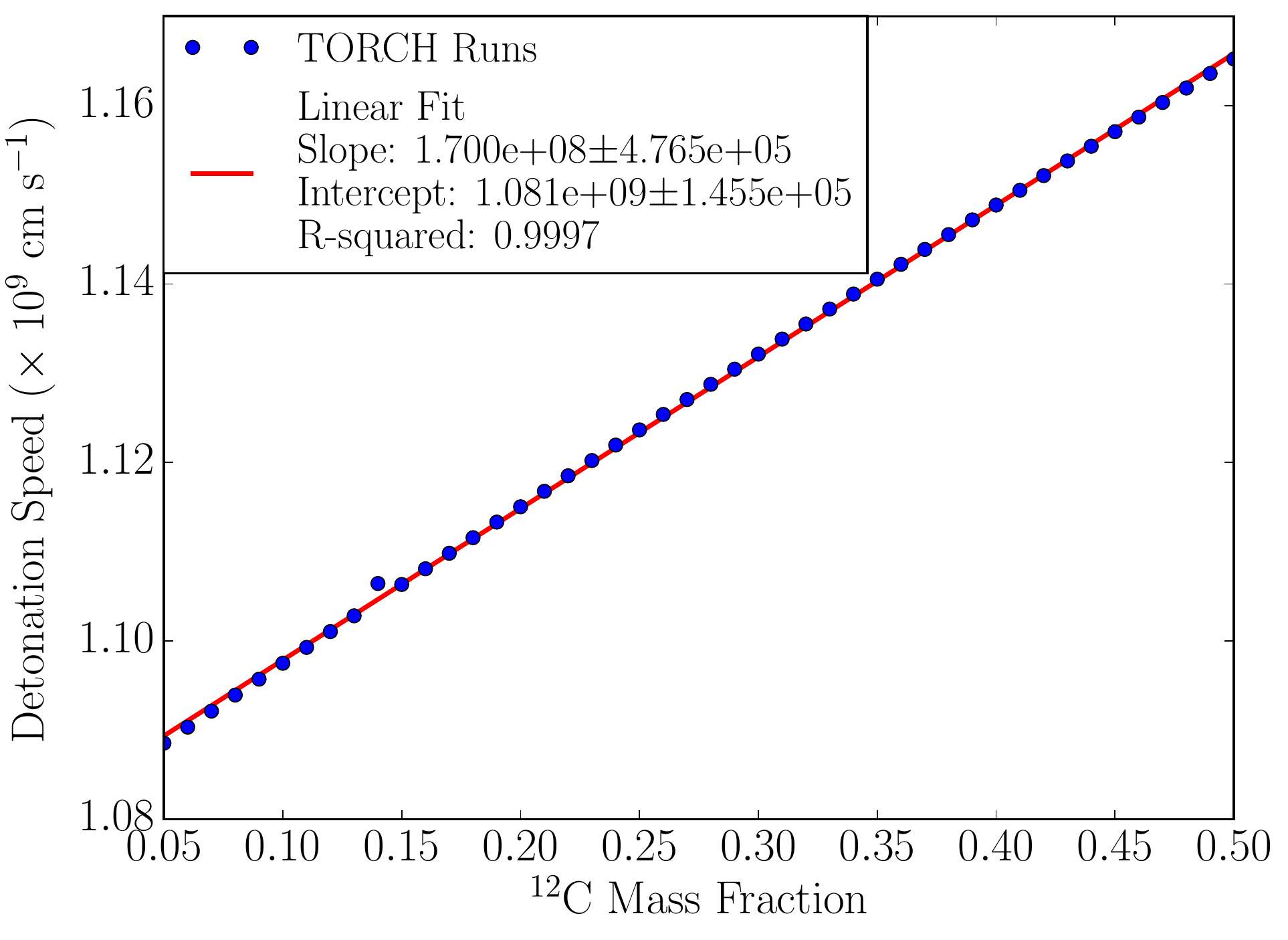}
	\caption{\label{fig:detonation_velocity_wfit} Self-supported detonation speed plotted versus carbon fraction in fuel with initial (pre-shock) density of $10^7$~\grampercc and initial composition of $X_{^{16}\rm O} = 0.48$ and $X_{^{22}\rm Ne} = 0.02$. The sum $X_{^{12}\rm C} + X_{^{20}\rm Ne}$ is kept constant.\\}
\end{figure}

\begin{figure}[t]
	\includegraphics[width=0.95\linewidth]{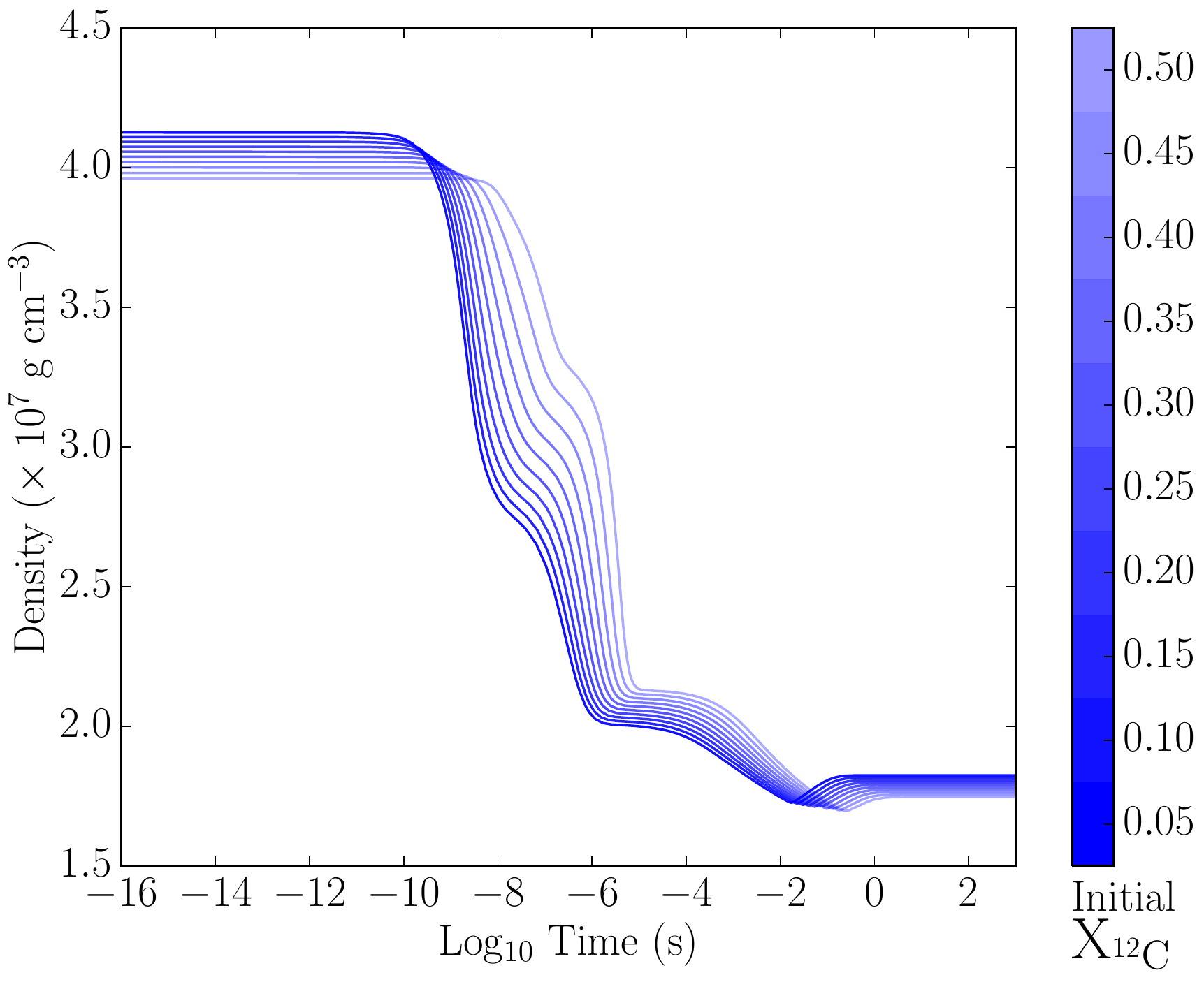}
        \caption{\label{fig:co_znd_multi_den}} Density structures shaded with varying $^{12}$C mass fraction, keeping constant the sum $X_{^{12}\rm C} + X_{^{20}\rm Ne}$. The rest of the composition is $X_{^{16}\rm O} = 0.48$ and $X_{^{22}\rm Ne} = 0.02$.
\end{figure}

The eigenvalue detonation speeds from ZND calculations in material with \C{12} fraction varying from 0.05 to 0.5 are shown in Figure \ref{fig:detonation_velocity_wfit}, demonstrating that self-supported detonations in this progenitor are feasible with only small variation in speed across this range of \C{12} fractions. 
Figure \ref{fig:co_znd_multi_den} shows
the effect of simultaneously adding \Ne{20} and reducing \C{12} on the density profile.
Lowering the \C{12} fraction weakens the shock and
lengthens the timescales of the step features, corresponding to more slowly
burning fuel and ash, as might be expected from the lower energy release
afforded by the \Ne{20}. However, it is noteworthy that no qualitatively
new features arise from the change in fuel source that would suggest
that more than 4 representative burning stages are needed.

\begin{figure}[t]
	\begin{minipage}{0.5\textwidth}
		\includegraphics[width=0.86\textwidth]{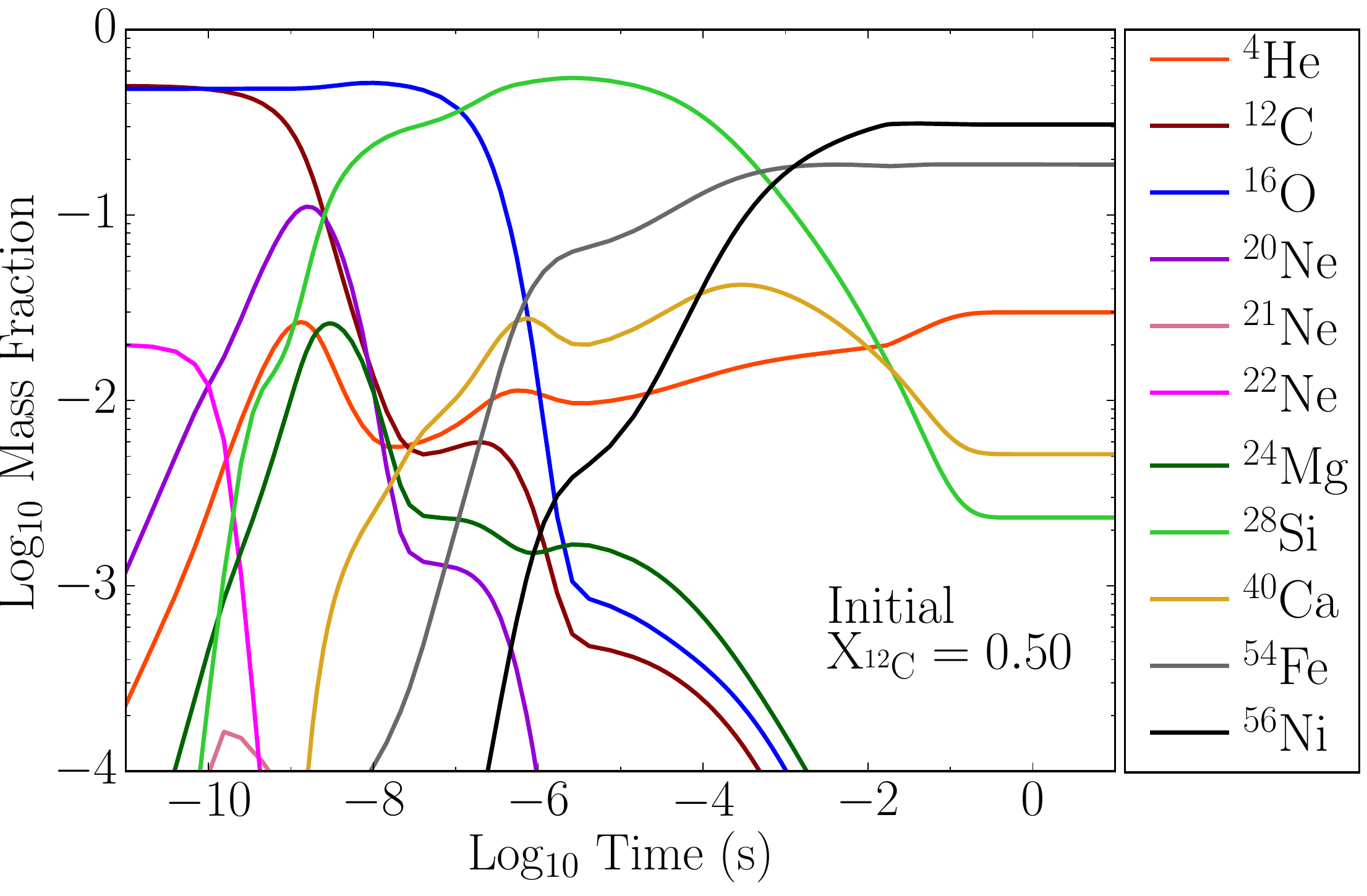}
	\end{minipage}	
	\hfill
	\begin{minipage}{0.5\textwidth}
		\includegraphics[width=0.86\textwidth]{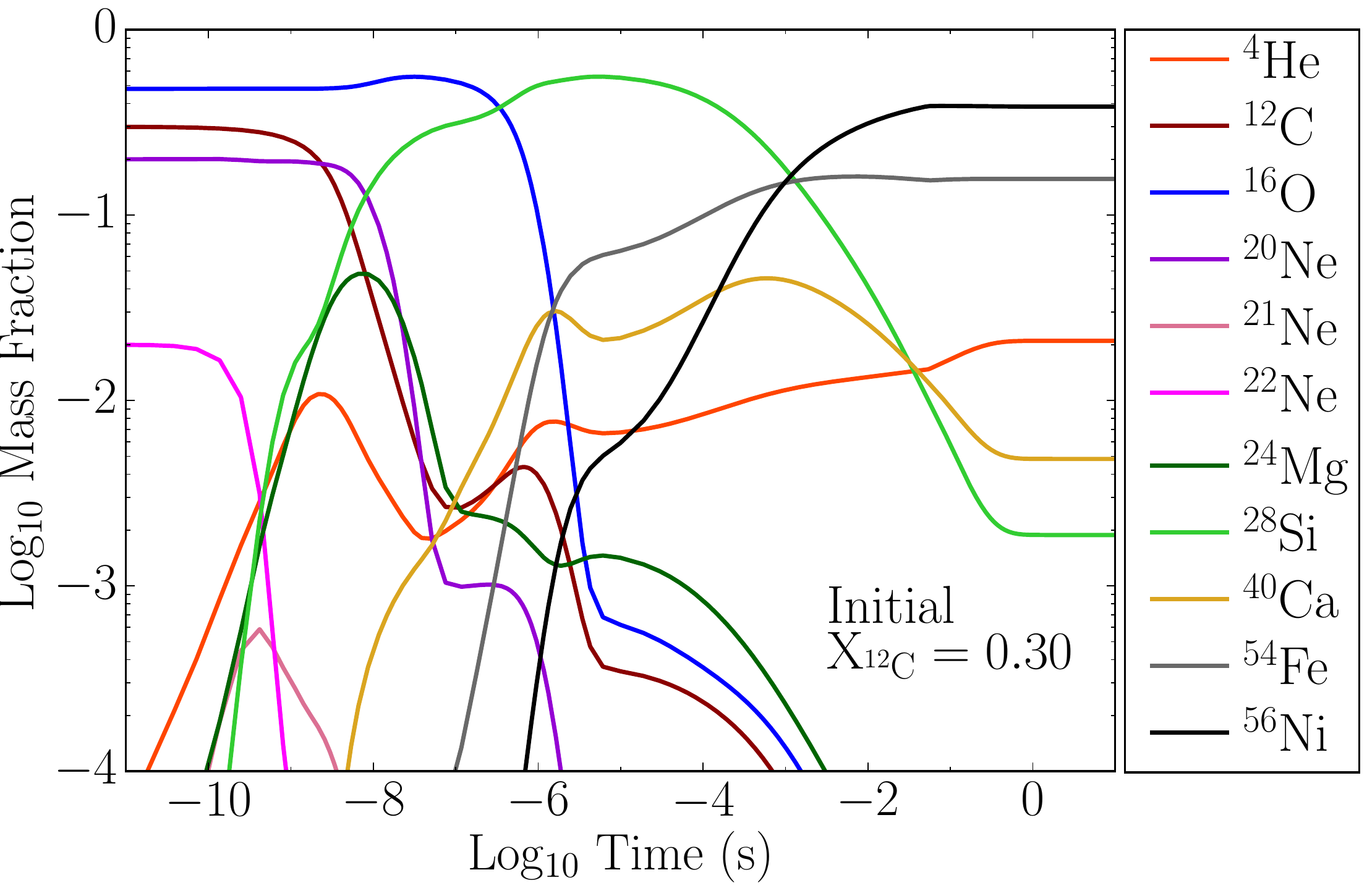}
	\end{minipage}
	\hfill
	\begin{minipage}{0.5\textwidth}
		\includegraphics[width=0.86\textwidth]{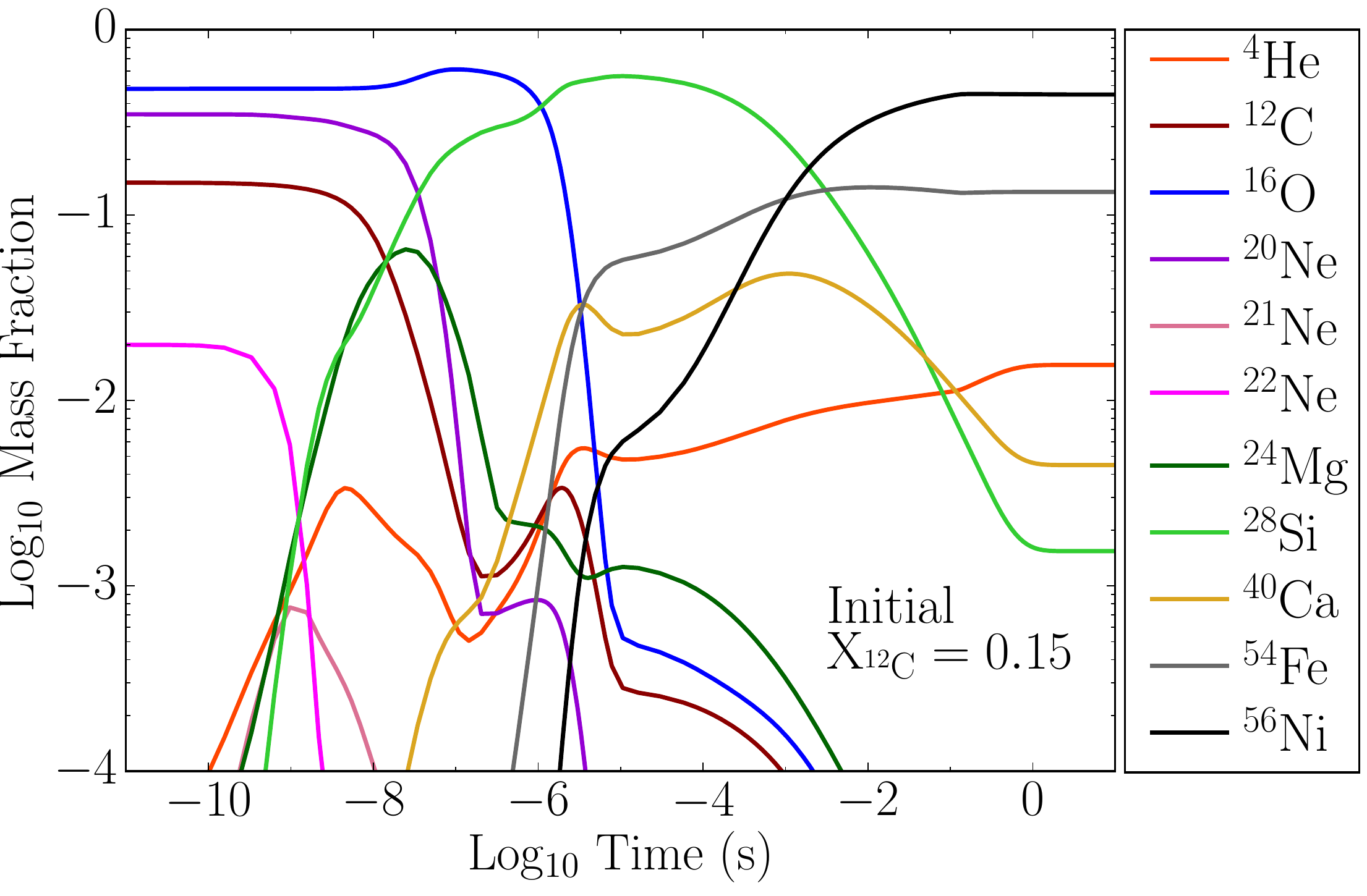}
	\end{minipage}
	\hfill
	\begin{minipage}{0.5\textwidth}
		\includegraphics[width=0.86\textwidth]{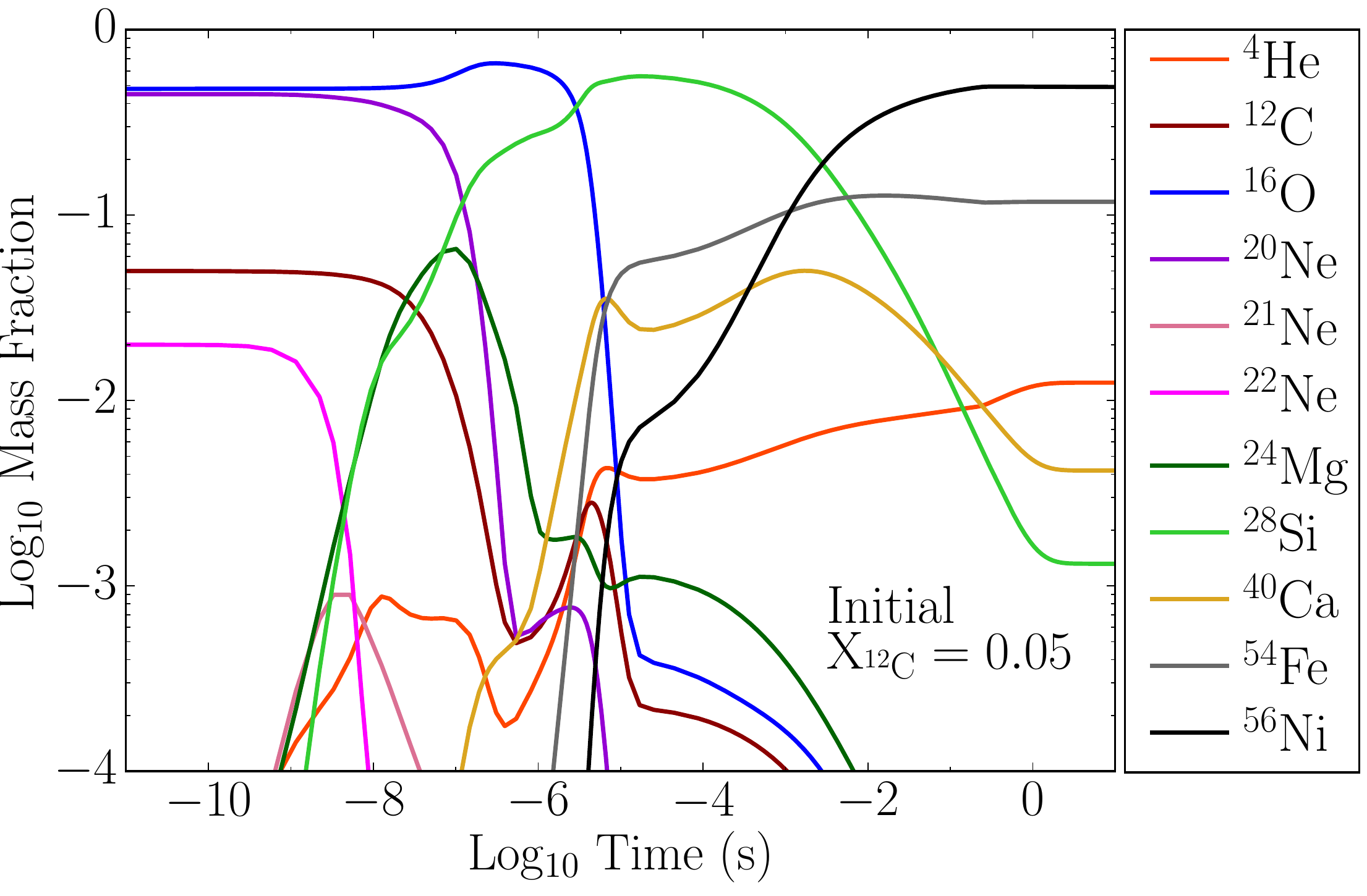}
	\end{minipage}
        \caption{\label{fig:znd_abundances} Mass Fraction Evolution for ZND Detonations with varying initial \C{12} and \Ne{20} mass fractions, calculated for an initial density of $10^7~\grampercc$.}

\end{figure}

Qualitative similarity between C-O and C-O-Ne detonation structures
are visible in the mass fractions as well.
To demonstrate this, the abundance structures with initial \C{12} abundances of 0.5, 0.3,
0.15, and 0.05 are shown in \figref{fig:znd_abundances}. 
We find that the \Ne{20} burns simultaneously with whatever \C{12} is present,
producing primarily \Si{28}.  The Ne-C burning stage is then followed by \Ox{16}
burning to silicon-group NSQE elements and then on to NSE,
just as in a model with no initial \Ne{20}, except for the progressively
later \Ox{16} burning time.

\begin{figure}[t]
	\includegraphics[width=1\linewidth]{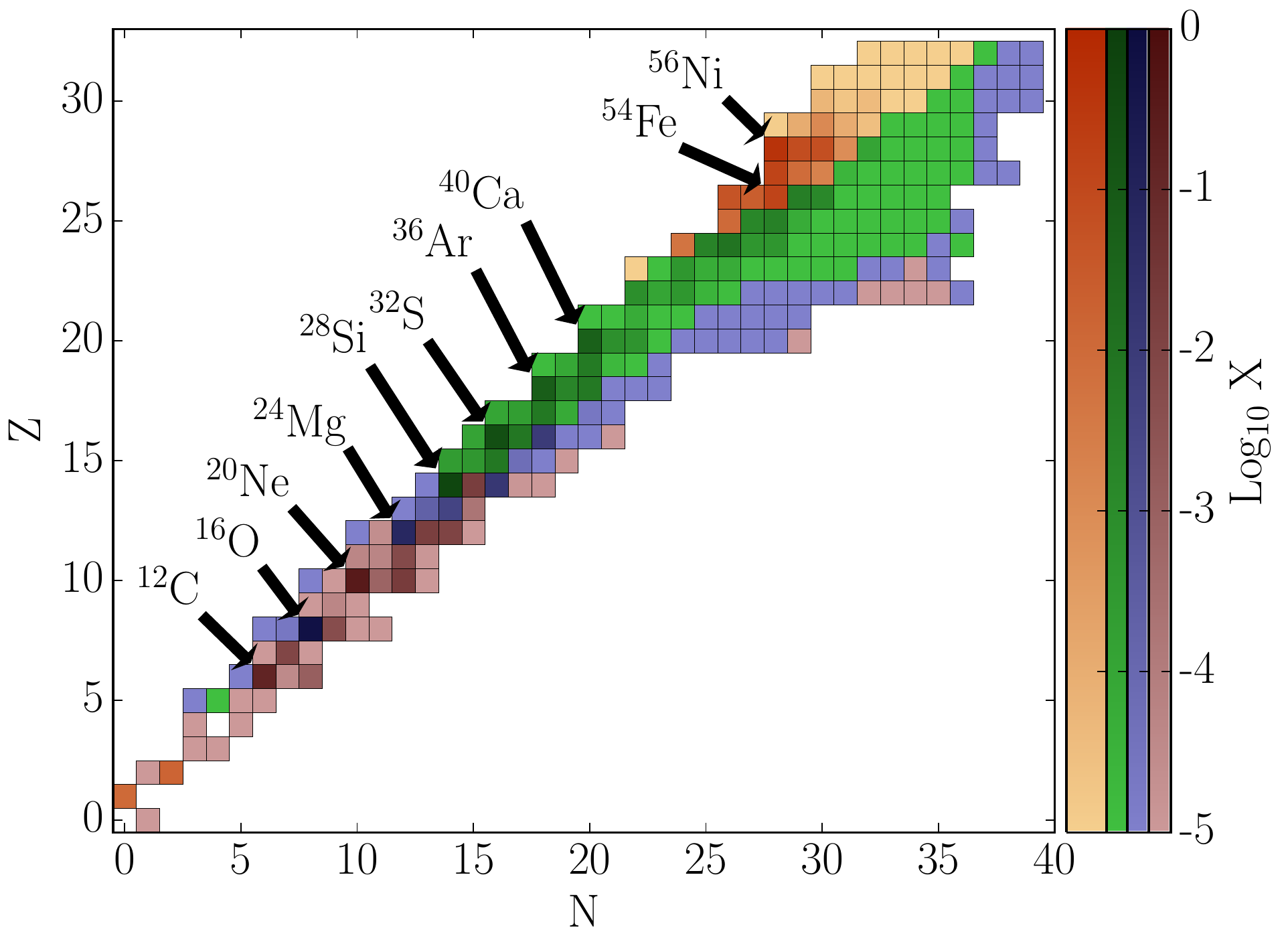}
	\caption{\label{fig:nuclides_xmax} Our reaction network of 225 nuclides is shown, with nuclides shaded by their maximum abundance and categorized based on whether they were maximally abundant before the earliest of the following crossing times: \C{12}-\Si{28} (purple), \Ox{16}-\Si{28} (blue), \Si{28}-\Fe{54} (green). Nuclides maximally abundant after the \Si{28}-\Fe{54} crossing time are shown in orange.
		The initial composition used 
		is $X_{^{16}O} = 0.48$, $X_{^{22}Ne} = 0.02$, $X_{^{12}C} = 0.15$, and $X_{^{20}Ne} = 0.35$.
	}
\end{figure}

A graphic representation of the most significant nuclides by mass
fraction and the stage in which they are important is shown in
\figref{fig:nuclides_xmax} for an initial \C{12} fraction of 0.15,
representative of the majority of the hybrid stellar profile
(cf. \figref{fig:progenitor_abundances}). Nuclides are categorized
based on the time at which they were maximally abundant in the network
and shaded by their maximum abundance.  Nuclides within the purple
color palette were maximally abundant during the initial
``fuel-burning'' stage after the beginning of fusion and before the
\Si{28} becomes equally abundant with \C{12}, the \C{12}-\Si{28}
crossing time $t_{\rm fa}$.  Blue nuclides were maximally abundant
after $t_{\rm fa}$ but prior to the \Ox{16}-\Si{28} crossing time
$t_{\rm aq}$, and nuclides maximally abundant after $t_{\rm aq}$ but
before the \Si{28}-\Fe{54} crossing time $t_{\rm qn}$ are plotted in
green. Nuclides maximally abundant following $t_{\rm qn}$ are IGEs
together with protons and alpha particles and are shaded in
orange. Note that for the time scales we adopt the notation of
\citep{townetal15}, which describes the burning stages in detail.

The burning model we are using computes the rates for progression
through the burning stages from the local temperature and the energy
release from a set of major fuel abundances, previously including
\C{12}, \Ox{16}, and \Ne{22}. We have found here that any \Ne{20} is
consumed along with \C{12} and that otherwise the burning is quite
similar to that with just \C{12} and \Ox{16} as principal
constituents. In consequence, the only necessary modification to our
burning model is to include \Ne{20} in the abundances of the initial
state in the burning model. This accounts for the difference in
binding energy of \Ne{20} compared to \C{12} and gives lower burning
temperatures. Additionally, throughout the majority of the progenitor
outside the core ignition region, due to prior convective mixing, the
\C{12} content is high enough that we will extrapolate the laminar flame speeds of
\citet{timmes92,Chametal08}. We consider this a reasonable
approximation since much of the flame propagation is dominated by
Rayleigh-Taylor overturn and turbulence.

\subsection{Mapping a \MESA\ Profile to \FLASH\ While Preserving Hydrostatic Equilibrium}
\label{subsec:mesa_flash_mapping}
The temperature and composition at the base of the convective zone in
the hybrid C-O-Ne white dwarf provides a natural flame initialization
region for simulation of thermonuclear runaway in the DDT scenario, so
we map the \MESA\ profile into the \FLASH\ domain preserving its
features at $4$~km spatial resolution. We do this by first converting
the \MESA\ model to a uniform grid by mass-weighted averaging of
quantities in \MESA\ zones with spacing less than $4$~km and using
quadratic interpolation to estimate quantities where \MESA\ zones have
spacing greater than $4$~km. Although our combustion model in
\FLASH\ does not evolve nuclide abundances, it uses the initial
abundances of \C{12}, \Ne{20} and \Ne{22} (assuming the rest is \Ox{16}) to compute the
initial mean nuclear binding energy and electron fraction.
Therefore, we also represent the full set of nuclides in the
\MESA\ profile by this reduced set of four nuclides in the uniformly
gridded profile, requiring the carbon mass fractions to be identical
because there is still sufficient \C{12} in the star to sustain a
detonation front. In addition, we use \Ne{22} in the reduced set to
account for the $Y_e$ of the full set of nuclides, and we constrain
\Ne{20} and \Ox{16} to be in the same ratio $R$ in both sets of
abundances. These constraints provide the following definitions for
the reduced abundances used for \FLASH:


\begin{subequations}
\begin{align}
X_{\C{12}}^{FLASH} &= X_{\C{12}}^{MESA} \\
X_{\Ne{22}}^{FLASH} &= 22 \cdot \left(\dfrac{1}{2}-Y_e^{MESA}\right) \\
R &= X_{\Ne{20}}^{MESA}/X_{\Ox{16}}^{MESA} \\
X_{\Ox{16}}^{FLASH} &= \frac{1 - X_{\C{12}}^{FLASH} - X_{\Ne{22}}^{FLASH}}{R + 1} \\
X_{\Ne{20}}^{FLASH} &= R \cdot X_{\Ox{16}}^{FLASH}.
\end{align}
\end{subequations}

Nothing constrains the resulting uniformly gridded profile to be in
hydrostatic equilibrium (HSE), however, so we next construct an
equilibrium profile by applying the HSE pressure
constraint
\begin{equation}\label{eq:hse_pressure}
  P_{EOS}(\rho_i,T_i,X_i) = \frac{g \Delta r}{2} (\rho_i +
  \rho_{i-1})\ .
\end{equation}
Starting at the uniformly gridded profile's central density and using
its temperature and composition in each zone together with the HELM
equation of state (EOS) of~\citet{castro1}, we solve
\eqqref{eq:hse_pressure} for the density in each zone. In
\eqqref{eq:hse_pressure}, $\Delta r$ indicates the zone width, $T_i$
is the temperature in zone $i$, $X_i$ is the composition in zone $i$,
and $g$ is the gravitational acceleration at the boundary of zone $i$
and $i-1$ due to the mass enclosed by zone $i-1$ and below. The
resulting uniformly gridded, equilibrium profile (\FLASH) and the
original profile (\MESA) are shown for comparison in Figures
\ref{fig:progenitor_abundances} and
\ref{fig:progenitor_temperatures}.
The values of the total mass before and after this
procedure differ by $6 \times 10^{-3}\ \mathrm{M_\odot}$.
This procedure produced a structure that was stable in \FLASH, with fluctuations in central density less than $3\%$, for at least $5$ seconds with no energy deposition.
Finally, we replaced the EOS
routine in the public \FLASH\ distribution with that
from~\citet{castro1} to obtain a more consistent tabulation.


\subsection{Combustion and Explosion Mechanisms In The \FLASH\ Code} 
\label{subsec:flash_flame}
To simulate the explosion from either hybrid or traditional
progenitor models, we use a modified version of \FLASH 
\footnote{\FLASH\ is available from
  \url{http://flash.uchicago.edu}. Our modifications for C-O WDs are
  available from \url{http://astronomy.ua.edu/townsley/code}.}, an
Eulerian adaptive-mesh compressible hydrodynamics code developed by
the ASC/Alliances Center for Astrophysical Thermonuclear Flashes at
the University of Chicago~\citep{Fryxetal00, calder.fryxell.ea:on}.
While \FLASH\ is capable of evolving thermonuclear reaction networks
coupled to the hydrodynamics, in order to treat a $ \le 1$ \cm\ flame
front in full-star simulations of Type Ia supernovae, we use a
coarsened flame model that uses an advection-diffusion-reaction (ADR)
scheme to evolve a scalar variable representing the progression of
burning from fuel to ash compositions as detailed
in~\citet{Caldetal07, townsley.calder.ea:flame,townetal2009}. An
additional scalar represents the burning progress from ash to
intermediate-mass silicon-group elements in nuclear statistical
quasi-equilibrium (NSQE). A final scalar represents the burning of
silicon-group elements to IGEs in nuclear statistical equilibrium
(NSE). The timescales for evolving these scalars are density and
temperature-dependent and are determined from self-heating and
steady-state detonation calculations with a $200+$ species nuclear
reaction network~\citep{Caldetal07,townetal15}. For our evaluation of
the suitability of this burning scheme for fuel with an admixture of
\Ne{20}, see \secref{subsec:one_burning_mods}.

\subsection{Deflagration-to-Detonation Transition}
\label{subsec:ddt_details}
This study presents simulations of the thermonuclear explosions of
both a hybrid C-O-Ne white dwarf
progenitor~\citep{denissenkovetal2015} and a C-O white dwarf
progenitor similar to that used in previously published suites of
\SNIa\ simulations~\citep{kruegetal12}. In both cases, we initialize
the simulations with a ``matchhead'' consisting of a region near or at
the white dwarf's center that is fully burned to nuclear statistical
equilibrium (NSE). The energy release from this initial burn ignites a
subsonic thermonuclear flame front that buoyantly rises and partially
consumes the star while the star expands in response. As detailed
in~\citet{townetal2009, jacketal2010}, in order to effect a
deflagration-to-detonation transition, we suppose the DDT point to be
parameterizable by a fuel density $\rho_{\mathrm{DDT}}$ at which the
subsonic flame reaches the distributed burning regime where the flame
region has become sufficiently turbulent that a supersonic detonation
front may arise, self-supported by the energy release from the nuclear
burning proceeding behind the detonation shock front.

We use a similar DDT parameterization for our C-O simulations as in
the \SNIa\ simulations of~\citet{kruegetal12}. Thus, when the
deflagration reaches a point where it is at $\rho_{\mathrm{DDT}} =
10^{7.2}~\grampercc$, we place a region fully burned to NSE
$32$~\km\ radially outwards from this point that is of size
$12$~\km\ in radius. Multiple DDT points may arise, but they are
constrained to be at least $200$~\km\ apart. Our DDT parameterization
for the hybrid C-O-Ne simulations is identical, and we chose to use a
DDT transition density of $\rho_{\mathrm{DDT}} = 10^{7.2}~\grampercc$
because it is the lowest density for which all the realizations in our
hybrid suite reliably reached the end of the detonation phase of the
explosion when nuclear burning progress freezes out.

\subsection{Mesh Geometry And Refinement}
\label{subsec:flash_geometry}
We performed our calculations in two-dimensional $\mathrm{z-r}$
cylindrical coordinates, extending radially from 0 to
$65,536$~\km\ and along the axis of symmetry from $-65,536$~\km\ to
$65,536$~\km. We selected a maximum refinement level corresponding to
$4$~\km\ resolution using the PARAMESH adaptive mesh refinement scheme
described in~\citet{Fryxetal00}. This resolution permitted efficiency
in performing many repeated simulations with different initial
conditions. The $4$~\km\ resolution was also informed by previous
resolution studies in~\citet{townsley.calder.ea:flame,townetal2009},
that found that in 2-D DDT simulations of C-O white dwarf explosions,
the trends with resolution of total mass above the DDT density
threshold at the DDT time are fairly robust. The amount of high
density mass at the DDT is important because it reflects the extent of
neutronization during the deflagration and thus correlates with the
IGE yield of the explosion. In addition, we wish to compare the hybrid
IGE yields and other explosion characteristics with those of
explosions from C-O white dwarf progenitors previously explored
in~\citet{kruegetal12}, which used $4$~\km\ resolution in \FLASH\ with
the same $\mathrm{z-r}$ geometry we describe above. We can thus make
our comparison robust by controlling for resolution and geometry
factors.

\section{Simulations and Results}
\subsection{Reference C-O WD Model}\label{refcomodel}
\begin{figure}[!ht]
	\includegraphics[width=\linewidth]{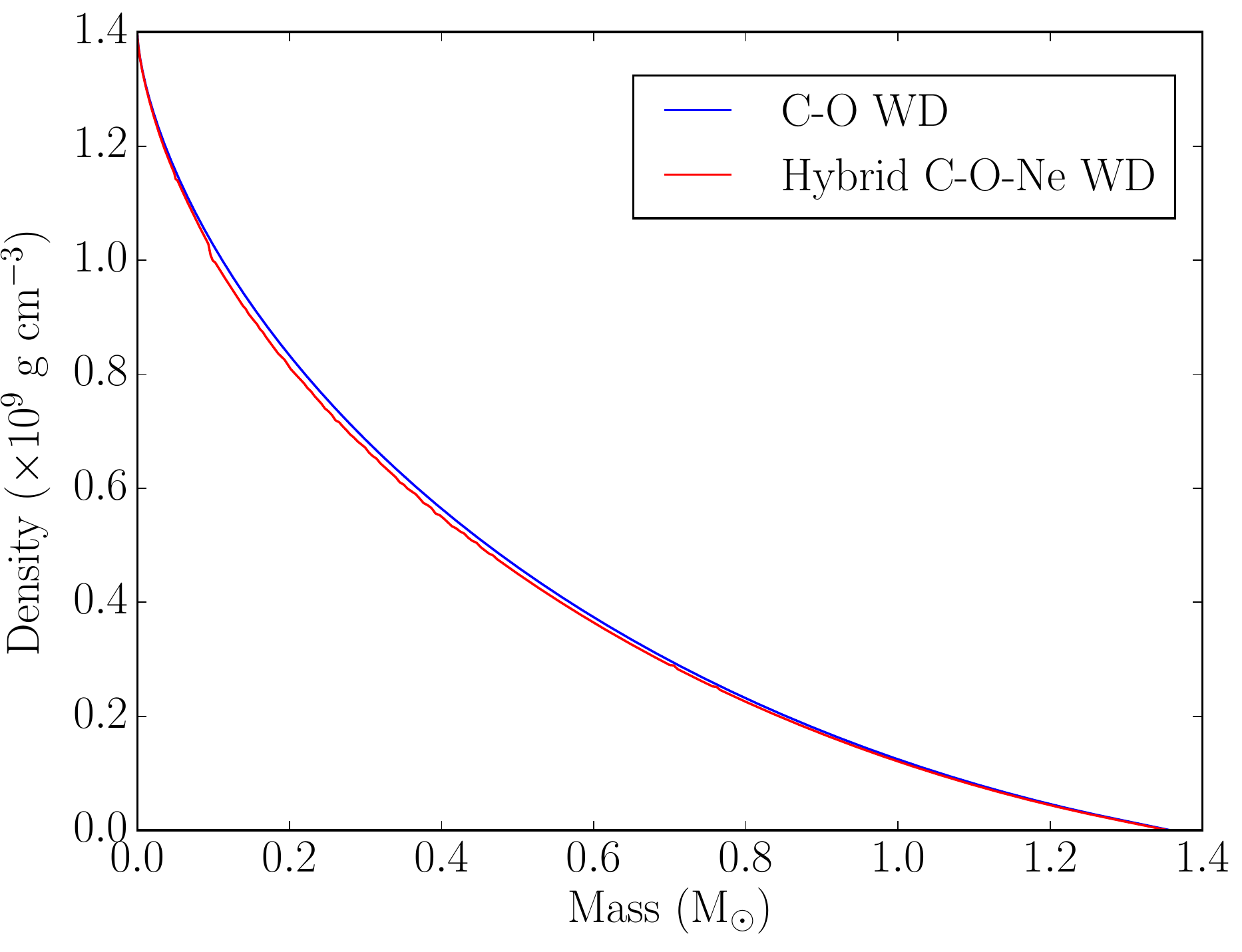}
	\caption{\label{fig:co_vs_cone_density} Density profiles of the hybrid C-O-Ne white dwarf (red) and the reference C-O white dwarf (blue), prepared at the same central density.}
\end{figure}

\begin{figure}[!ht]
	\includegraphics[width=\linewidth]{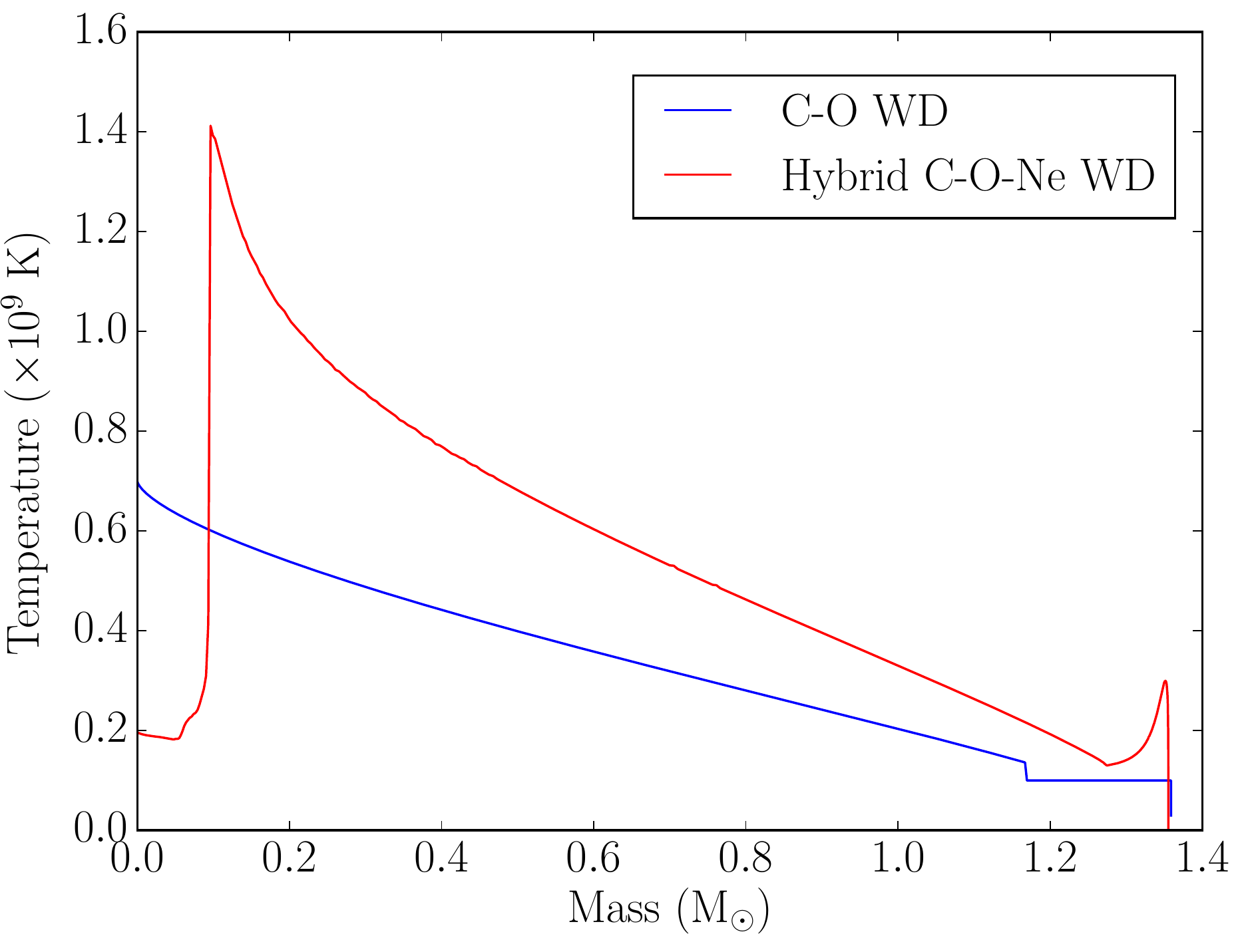}
	\caption{\label{fig:co_vs_cone_temp} Temperature profiles of the hybrid C-O-Ne white dwarf (red) and the reference C-O white dwarf (blue).}
\end{figure}

To compare the resulting features of Type Ia supernovae produced by
the hybrid model to those of previous studies of centrally ignited C-O
white dwarfs in the DDT
paradigm~\citep{Krueger2010On-Variations-o,kruegetal12}, we generate a
reference C-O model. The reference model has the same central density
as the hybrid model with a central temperature of $7 \times 10^8$~K, a
core composition of (\C{12} = 0.4, \Ne{20} = 0.03, \Ox{16} = 0.57) and
envelope composition of (\C{12} = 0.5, \Ne{20} = 0.02, \Ox{16} =
0.48). For comparison, Figures \ref{fig:co_vs_cone_density} and
\ref{fig:co_vs_cone_temp} show the density and temperature profiles of
the C-O and C-O-Ne hybrid WD.
Because the peak temperature of the C-O WD
is at the center, we initialize a suite of deflagrations in the C-O
model by instantaneously burning a region centered on the center of
the star with a nominal radius of $150$~km and 35 sets of randomly
seeded amplitude perturbations including those of~\citet{kruegetal12}, which
followed the method of~\citet{townetal2009}.

\subsection{Ignition Conditions for the C-O-Ne Hybrid WD}
\label{subsec:hybrid_ignition_parameters}
Given the temperature and composition profile of the hybrid model, if
it is to undergo thermonuclear runaway, \C{12} ignition will begin at
the base of the convective zone coinciding with the temperature peak
and \C{12} abundance of $\approx 0.14$. We therefore initialize a
deflagration by instantaneously burning a thin shell of material at
the peak temperature corresponding to a stellar radius of $350$~km.
Due to lack of constraints on the exact geometry of the ignition
region, we parameterize the thickness of the burned shell by an
angular sinusoidal function with variable harmonic number and
amplitude as shown in \figref{fig:cone_delayed_core} at 0.0~\second.
The harmonic number controls the number of initially burned regions
and the amplitude controls their size. With this form of
initialization, we generate a suite of 35 hybrid realizations
corresponding to a range of initially burned masses from $3.0\times
10^{-3}~\Msun$ to $1.3\times 10^{-2}~\Msun$ and a number of initial
burned regions from 1 to 10.

\begin{figure}[!ht]
	\includegraphics[width=\linewidth]{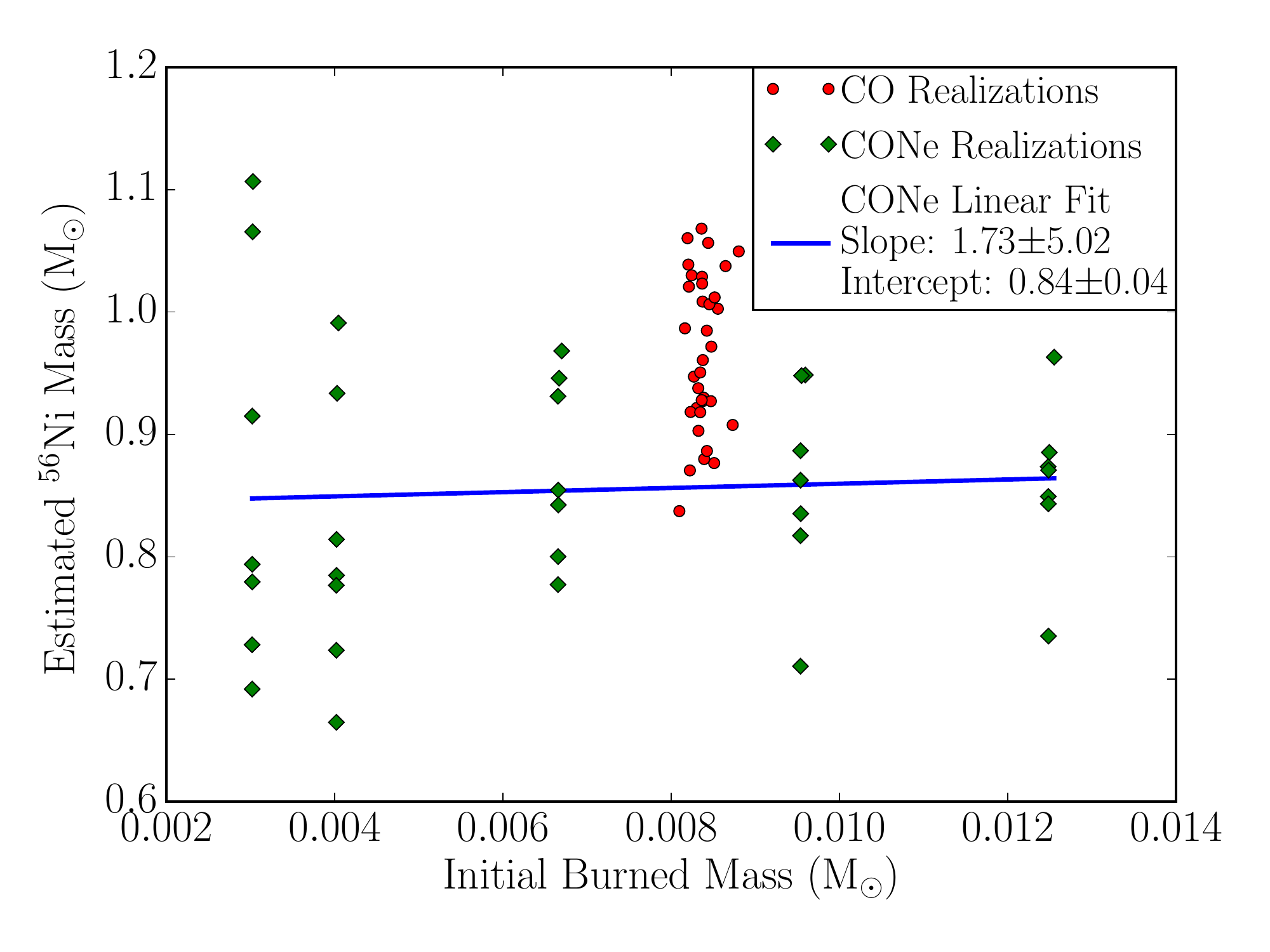}
	\caption{\label{fig:co_vs_cone_ignition_bm} Dependence of the final \Ni{56} yield on the initially burned mass at ignition for the 35 C-O (red) and 35 C-O-Ne realizations (green).}
\end{figure}

\begin{figure}[!ht]
	\includegraphics[width=\linewidth]{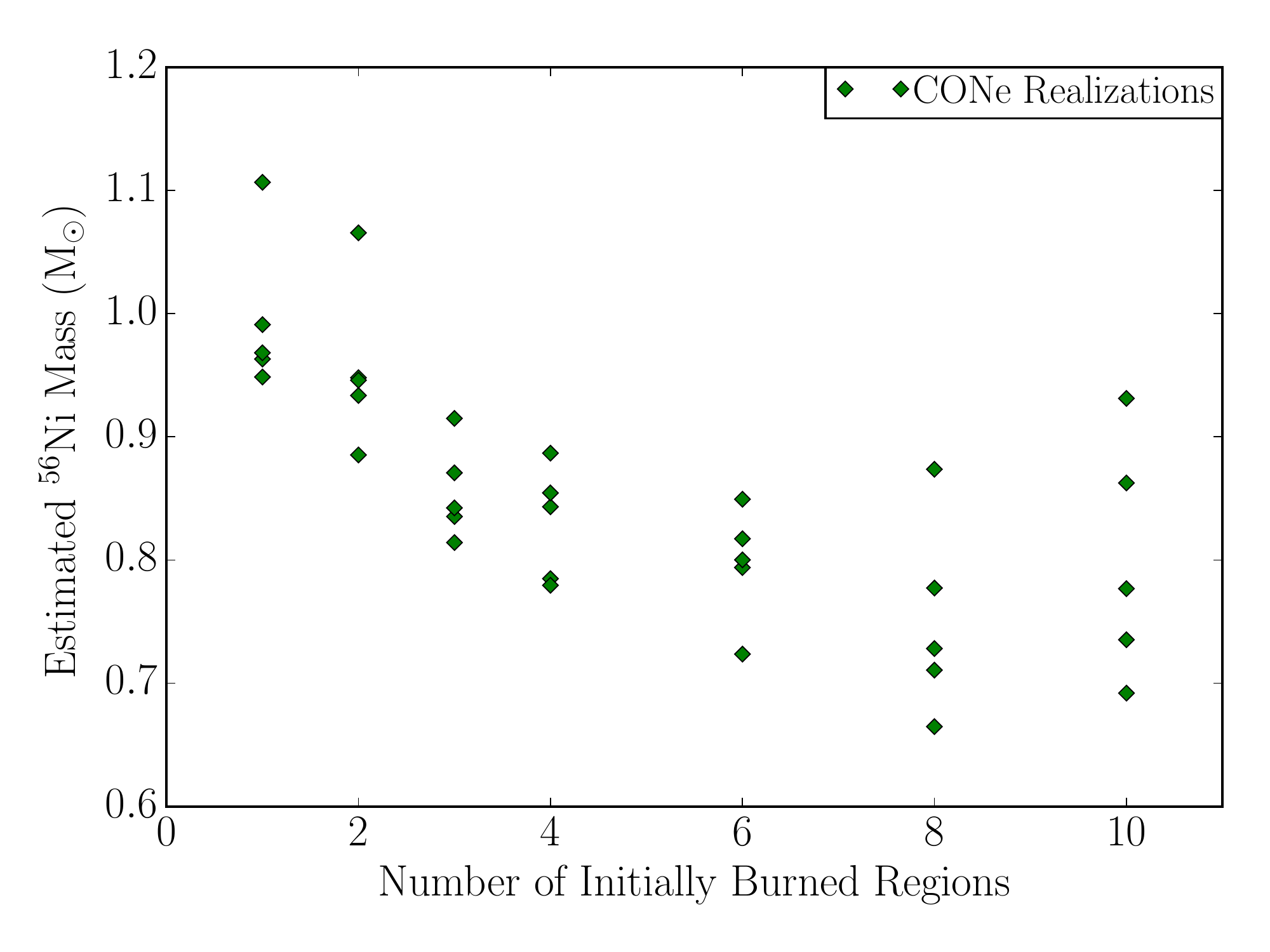}
	\caption{\label{fig:cone_ignition_mp} Dependence of the final \Ni{56} yield on the distribution of the initially burned mass at ignition for the 35 C-O-Ne realizations.}
\end{figure}

We demonstrate the influence of the number-and-amplitude parameters
using the final \Ni{56} yield as a proxy for the explosion results in
\figref{fig:co_vs_cone_ignition_bm} and \figref{fig:cone_ignition_mp}.
Because we use the same ignition geometry and conditions for the C-O
realizations as were used in~\citet{kruegetal12}, the general size of
the initially burned region for the C-O realizations remains very near
$0.0084~\Msun$. However, because the nature of the ignition in the
C-O-Ne hybrid progenitor is unknown, we chose to sample the
number-and-amplitude parameter space to provide a range of initially
burned masses for comparison. In spite of the scatter in
\figref{fig:co_vs_cone_ignition_bm}, we performed a linear fit between
the \Ni{56} yield and initially burned mass for the C-O-Ne
realizations, obtaining a slope indistinguishable from zero within
uncertainties and an intercept that matches the average \Ni{56} yield
for the C-O-Ne realizations (\tabref{YieldSummaryTable}). We also show
in \figref{fig:cone_ignition_mp} that most of the variation in \Ni{56}
yield from the C-O-Ne realizations originates from the interplay
between the number of ignition regions and their size. The more
ignition regions that are used, the greater effect the variation on
their size has on the spread in \Ni{56} yields.

Using these parameters to vary the initial burned mass and its
distribution within the progenitor, we evaluate the effect of this
parameterization on the estimated \Ni{56} yield, IGE yield, binding
energy, and other explosion properties in the following section.

\subsection{Characteristics of Explosions from C-O and C-O-Ne WD Models}
\begin{figure}[!ht]
\includegraphics[width=\linewidth]{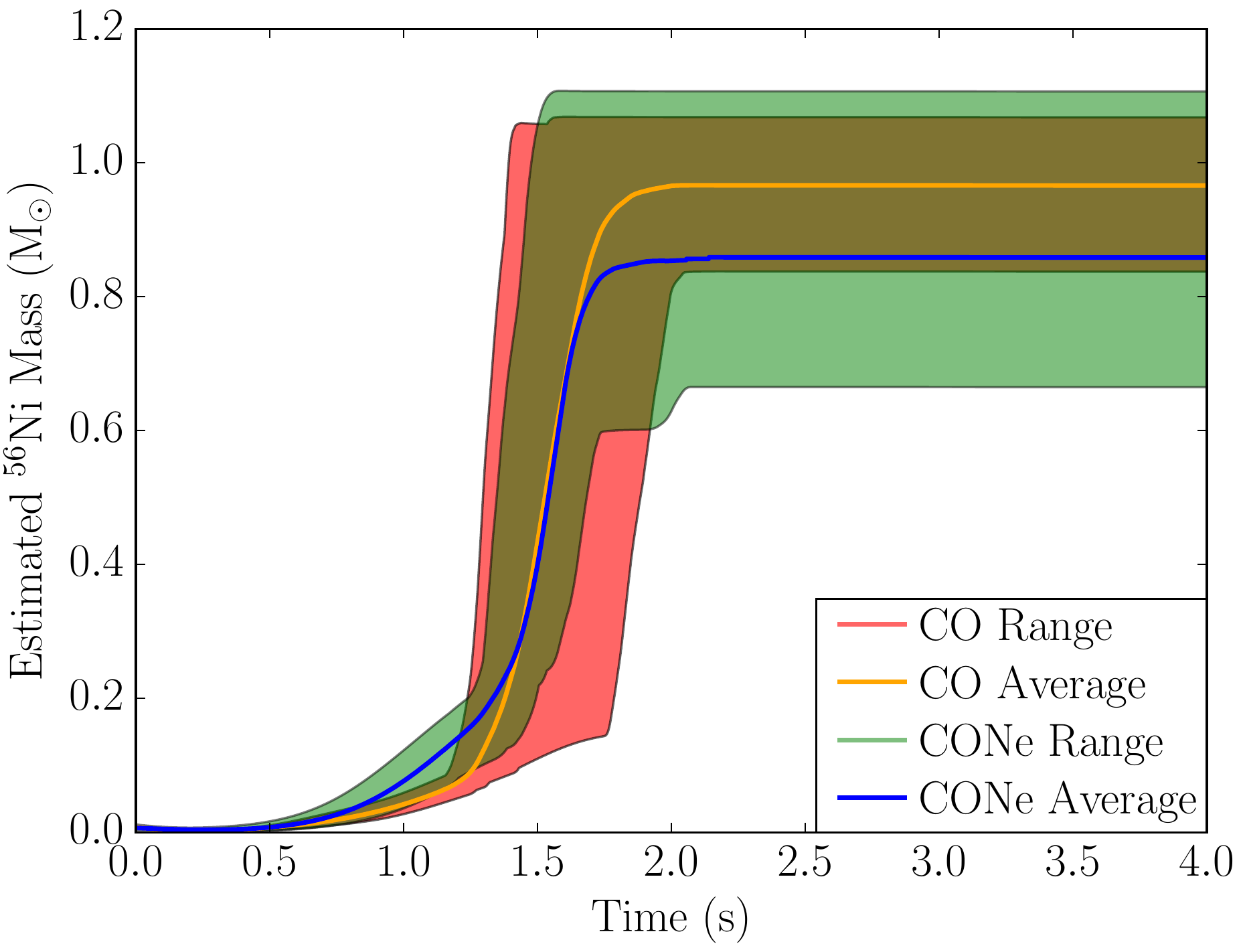}
\caption{\label{fig:est_ni_mass} Evolution of the estimated $^{56}$Ni yields for C-O and Hybrid C-O-Ne WD Realizations. The time-averaged value among C-O realizations is shown in yellow, with the full range of values at any point in time for the C-O realizations shown in red. Likewise, the time-averaged value among C-O-Ne realizations is shown in blue and their range of values shown in green.}
\end{figure}

\begin{figure}[!ht]
	\includegraphics[width=\linewidth]{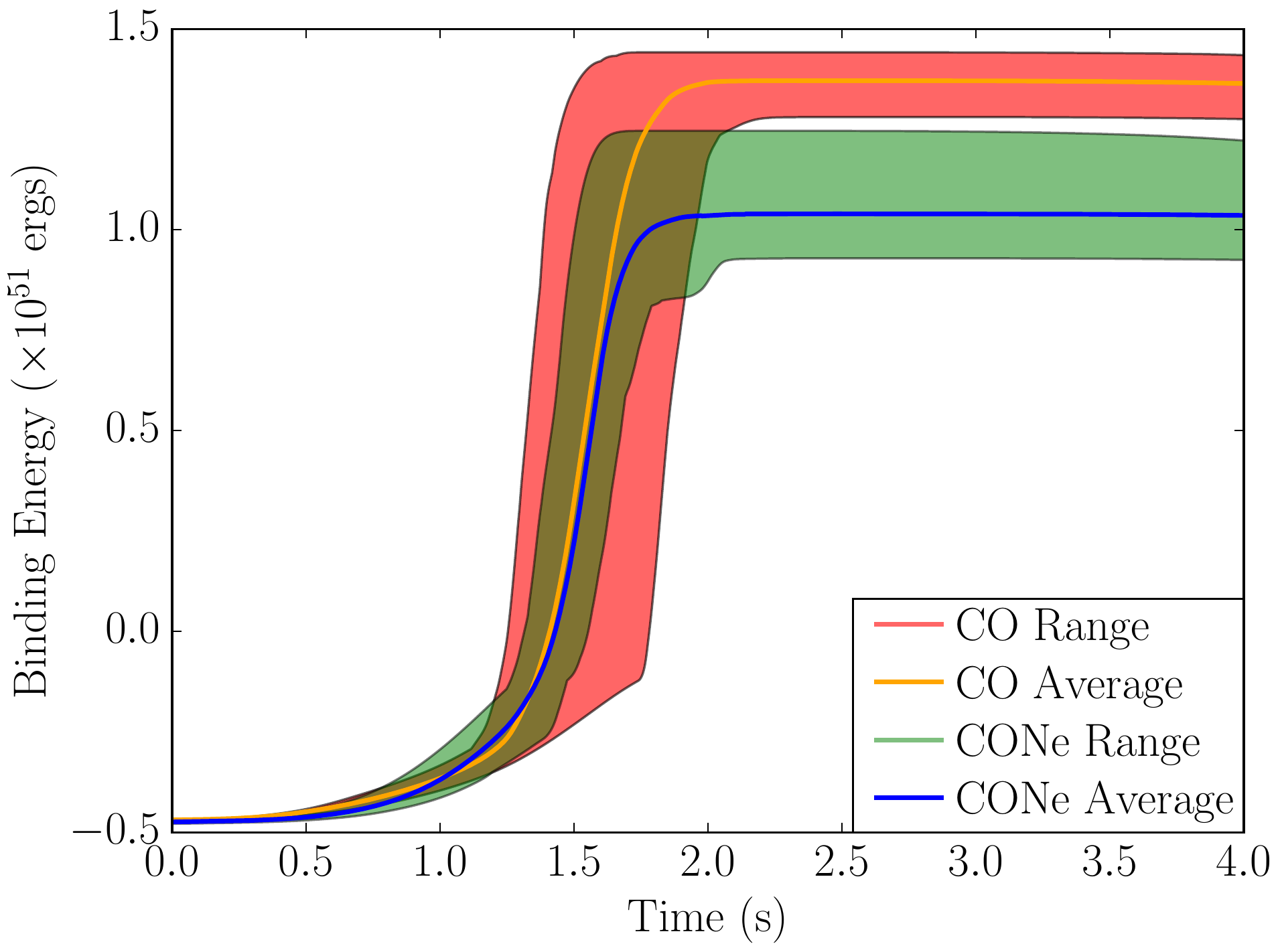}
	\caption{\label{fig:e_binding} Evolution of the binding energy for C-O and Hybrid C-O-Ne WD Realizations. The time-averaged value among C-O realizations is shown in yellow, with the full range of values at any point in time for the C-O realizations shown in red. Likewise, the time-averaged value among C-O-Ne realizations is shown in blue and their range of values shown in green.}
\end{figure}

We simulate the explosions of the hybrid and C-O realizations through
the end of the detonation phase and compare their features in Figures
\ref{fig:est_ni_mass}, \ref{fig:e_binding},
\ref{fig:mass_dens_lt_2e7}, \ref{fig:ige_vs_highrho_ddt},
\ref{fig:ni56_vs_nse_mass}, \ref{fig:ni56_nse_mass_ratio}, and
\ref{fig:nse_burn_mass} below. We compare the \Ni{56} yields in the
C-O and hybrid models, estimated from $Y_e$ and the NSE progress
variable, by assuming the composition in NSE is \Ni{56} plus equal
parts \Fe{54} and \Ni{58}, as described in
\citet{townetal2009,Meaketal09}.


\begin{figure*}[!ht]
  \begin{minipage}{0.24\textwidth}
    \includegraphics[width=\linewidth]{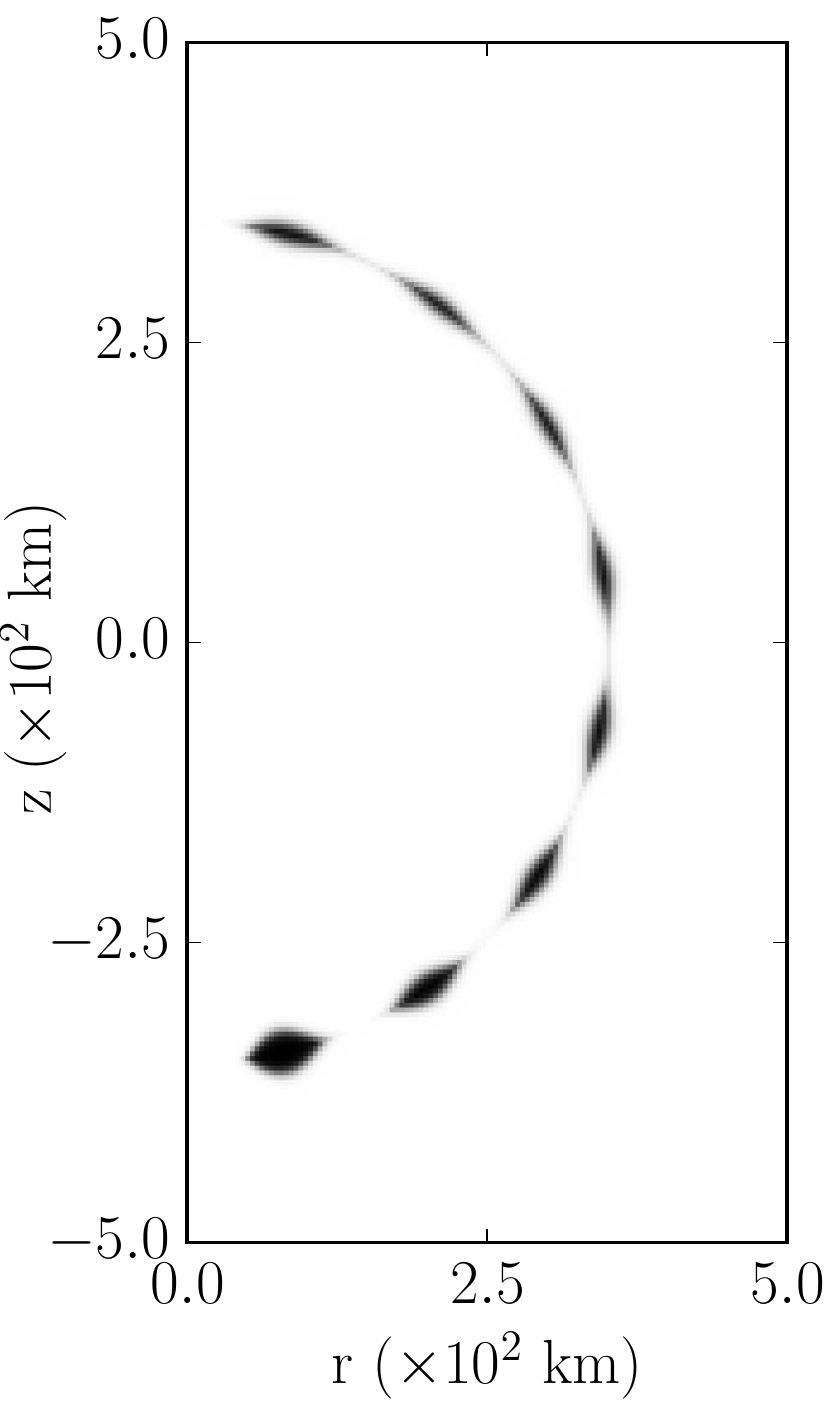}
  \end{minipage} \hfill
	\begin{minipage}{0.24\textwidth}
		\includegraphics[width=\linewidth]{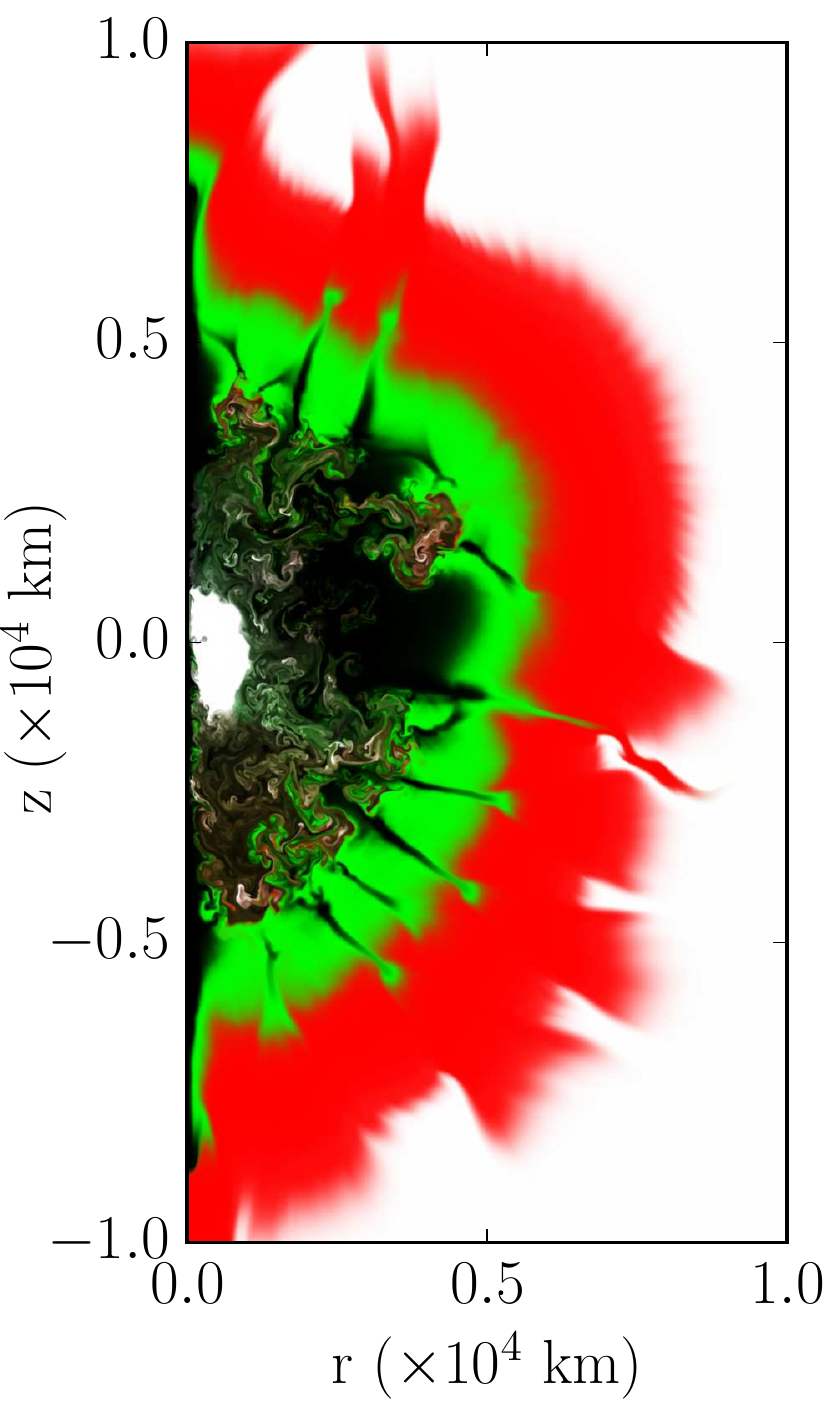}
	\end{minipage} \hfill 
        \begin{minipage}{0.24\textwidth}
		\includegraphics[width=\linewidth]{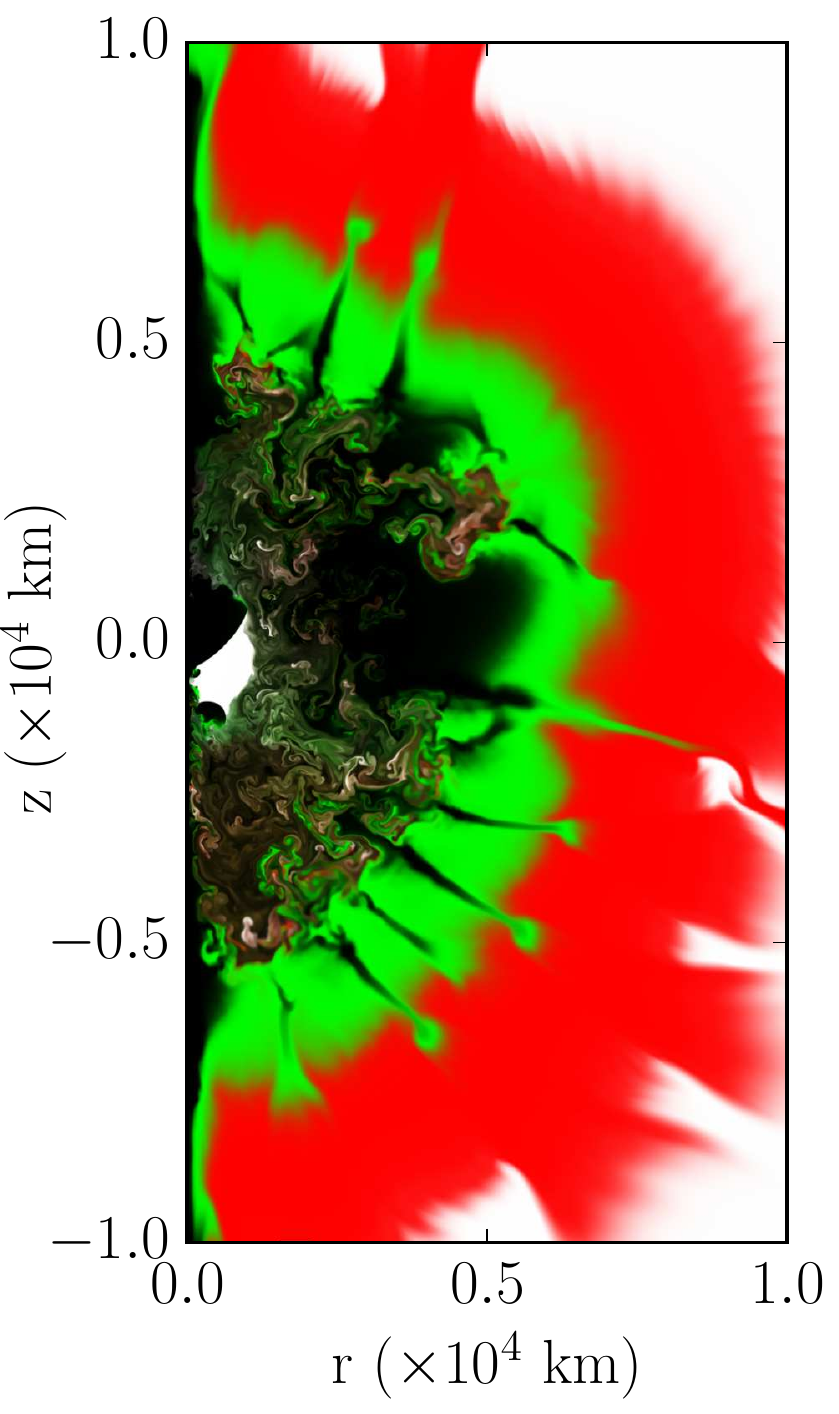}
        \end{minipage} \hfill 
        \begin{minipage}{0.24\textwidth}
		\includegraphics[width=\linewidth]{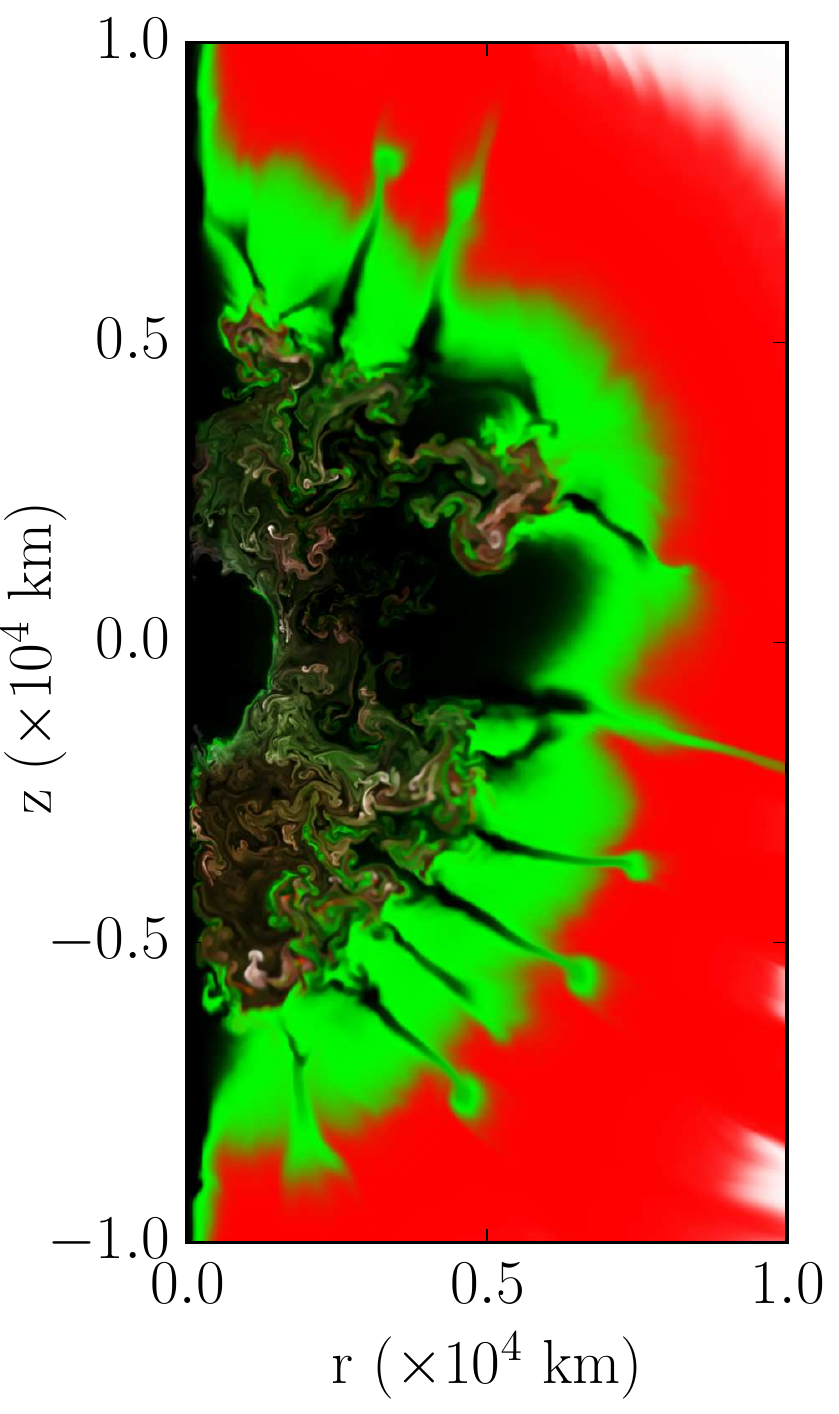}
              \end{minipage} \caption{\label{fig:cone_delayed_core} Progress of the burning front into the stellar core for one hybrid C-O-Ne realization, delayed relative to complete burning throughout the rest of the star. For reference, the initially burned geometry is shown at left. Material is shaded based on the reaction progress variables so that \textbf{White} denotes unburned fuel (\C{12}, \Ox{16} \& \Ne{20}) and \textbf{Red} denotes ash from \C{12} and \Ne{20}-burning. \textbf{Green} then denotes material in quasi-nuclear statistical equilibrium (primarily intermediate-mass silicon-group elements), and \textbf{Black} denotes material in nuclear statistical equilibrium (IGEs and $\alpha$-particles). From left to right, the burning is shown at 0.0~\second, 1.9~\second, 2.0~\second, and 2.1~\second. The DDT time for this realization is $\approx \mathrm{1.4}~\second$.}
\end{figure*}

\begin{figure}[!ht]
	\includegraphics[width=\linewidth]{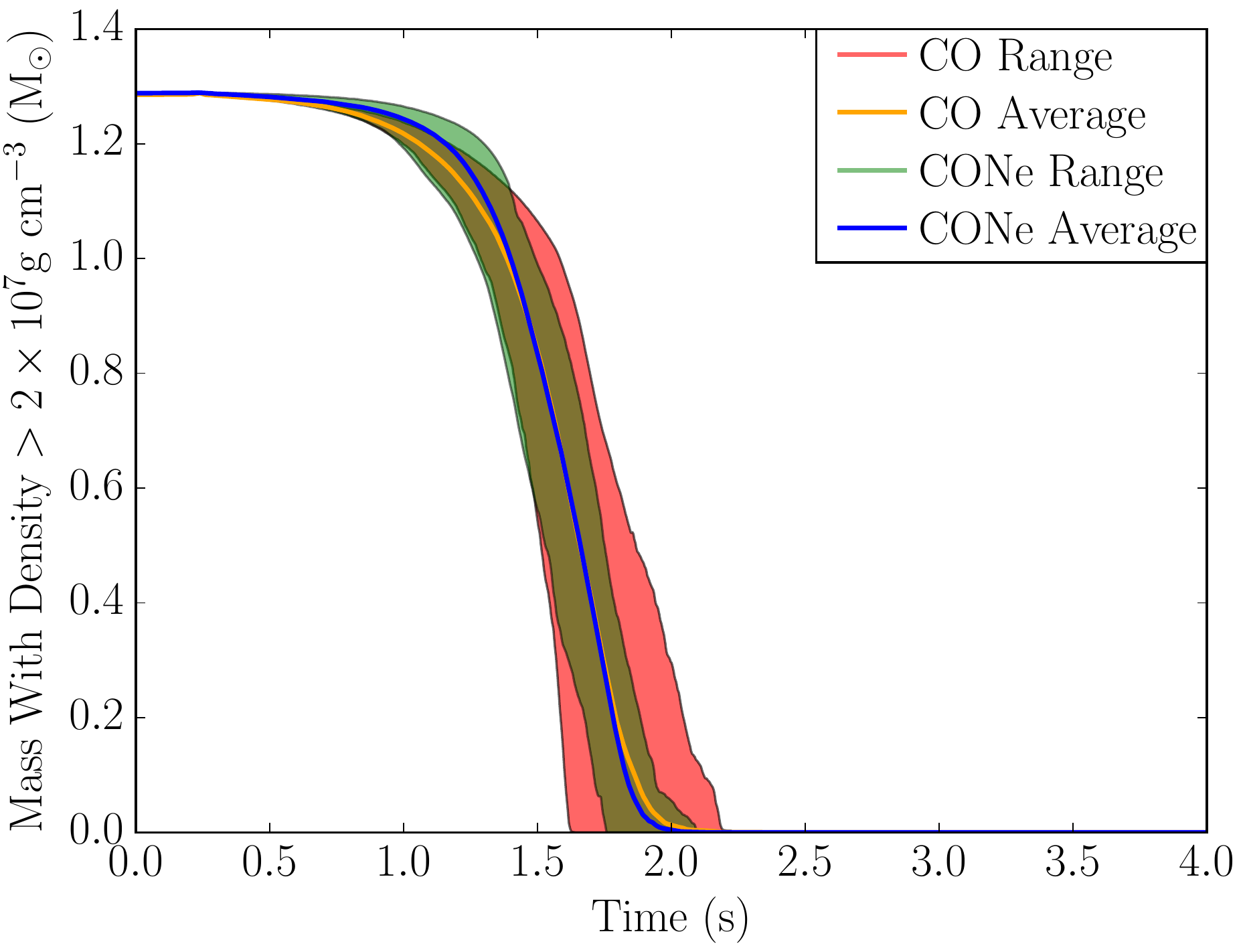}
	\caption{\label{fig:mass_dens_lt_2e7} Evolution of the total mass having density greater than $2 \times 10^7~\grampercc$ for C-O and Hybrid C-O-Ne WD Realizations. The time-averaged value among C-O realizations is shown in yellow, with the full range of values at any point in time for the C-O realizations shown in red. Likewise, the time-averaged value among C-O-Ne realizations is shown in blue and their range of values shown in green.}
\end{figure}

\begin{figure}[!ht]
	\includegraphics[width=\linewidth]{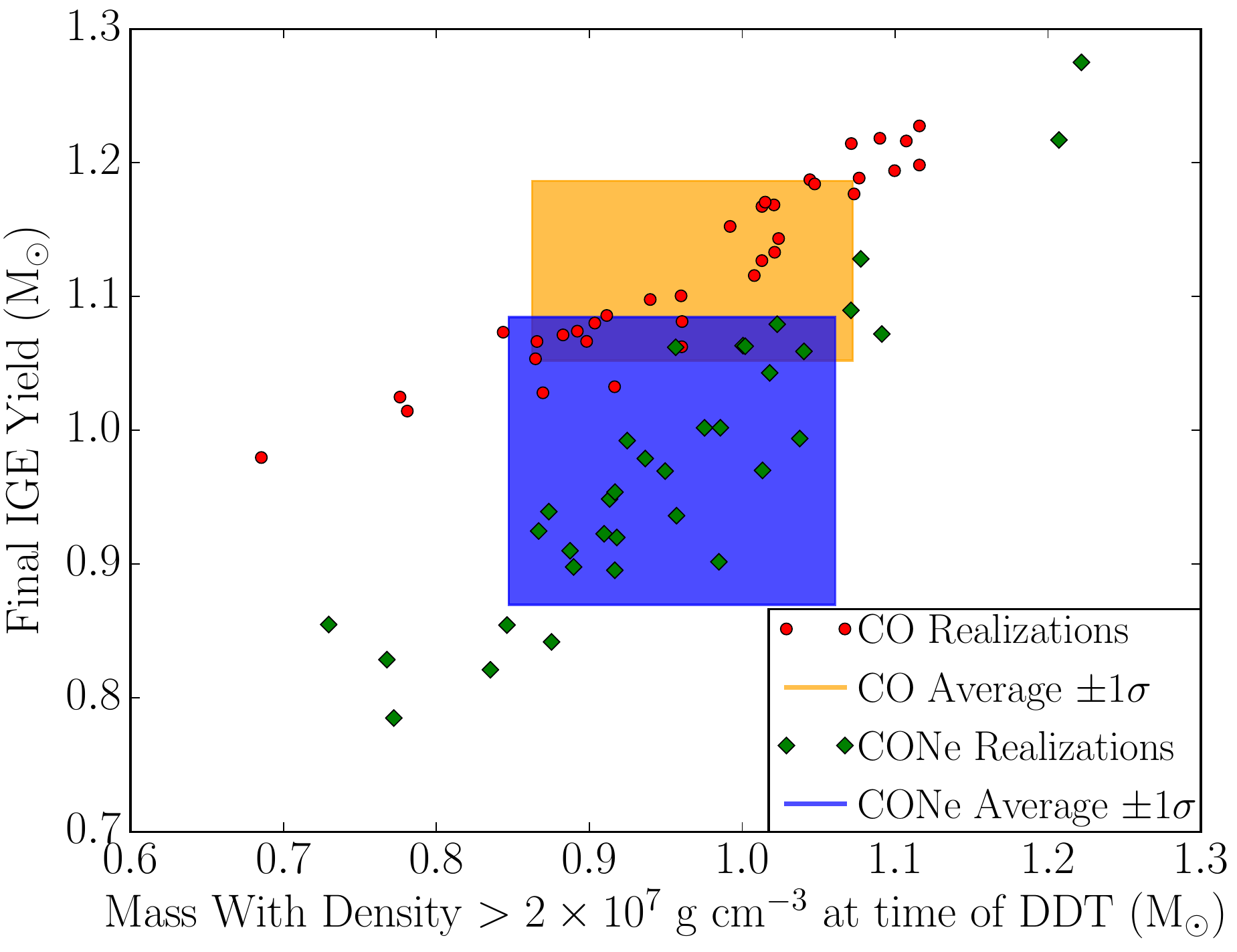}
	\caption{\label{fig:ige_vs_highrho_ddt} Final IGE yields for a range of deflagration expansion, where the degree of expansion is characterized by the mass denser than $2\times 10^7~\grampercc$ at the DDT time. Individual C-O (red) and C-O-Ne (green) realizations are shown, together with a rectangular area centered at their average values with size extending $\pm 1 \sigma$ along each axis for C-O (yellow) and C-O-Ne (blue) groups.}
\end{figure}

\begin{figure}[!ht]
	\includegraphics[width=\linewidth]{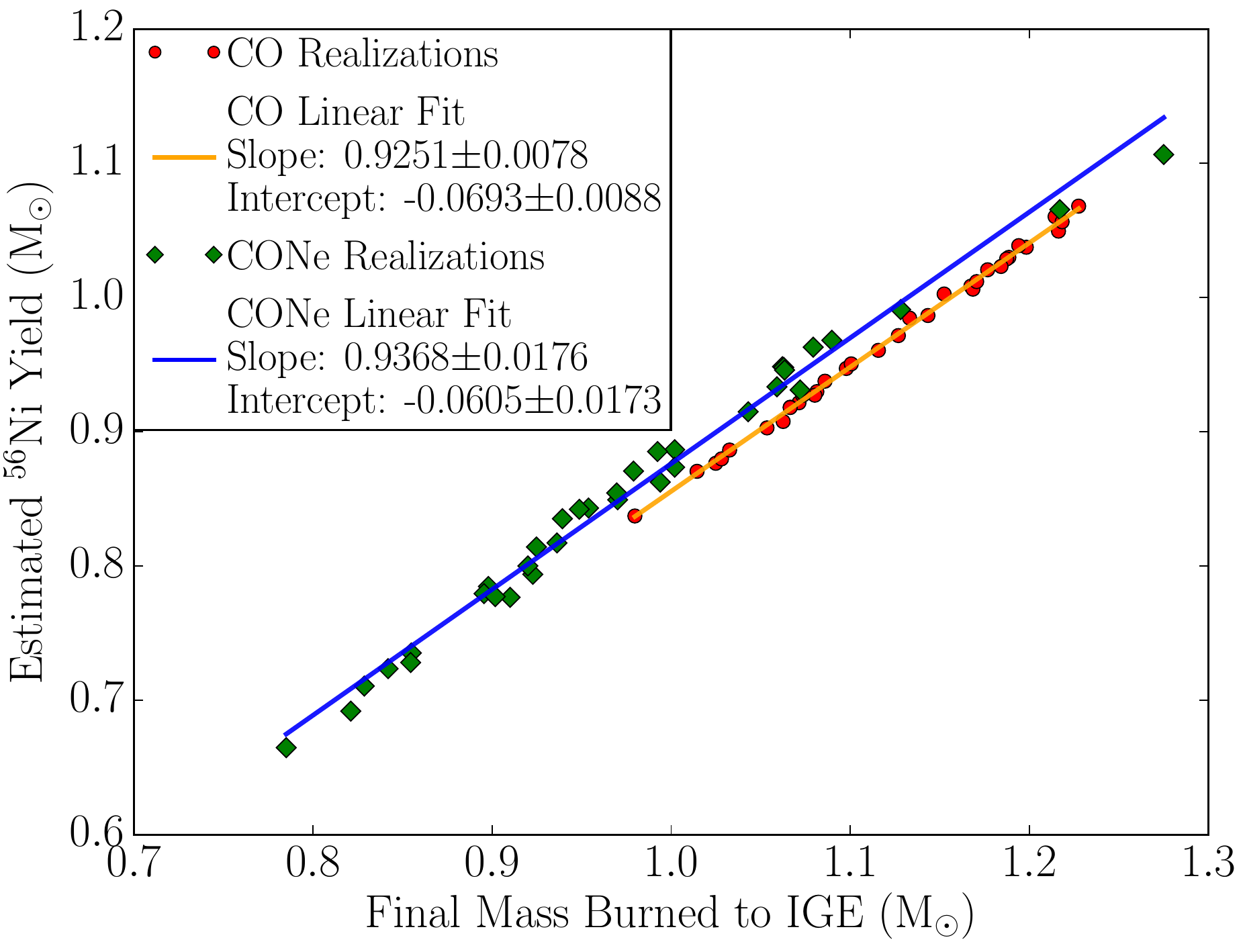}
	\caption{\label{fig:ni56_vs_nse_mass} Production of $^{56}$Ni and Mass Burned to IGE for C-O (red) and Hybrid C-O-Ne (green) WD Realizations. To estimate the overall fraction for each case, a linear fit is shown for C-O (yellow) and C-O-Ne (blue) realizations.}
\end{figure}

\begin{figure}[!ht]
	\includegraphics[width=\linewidth]{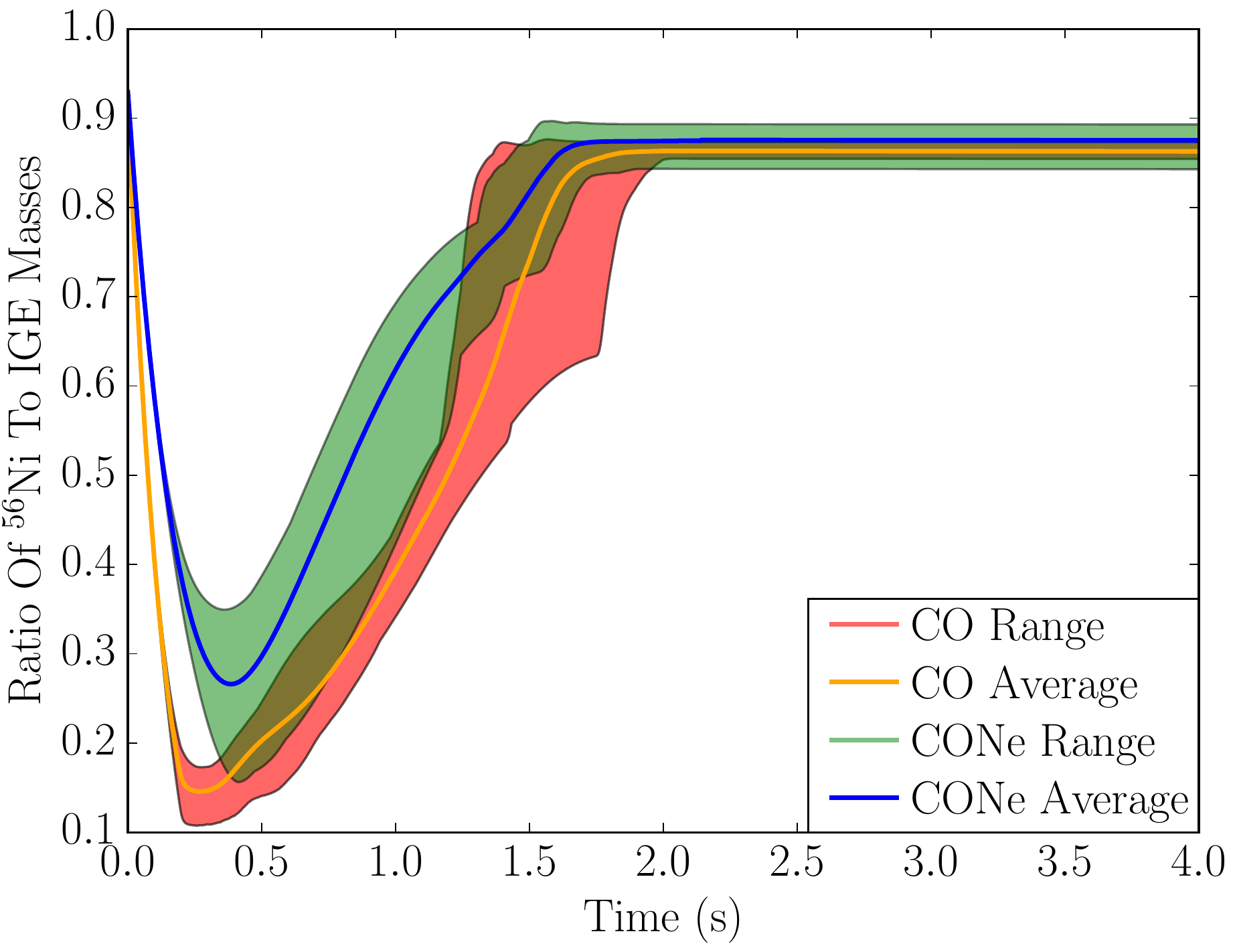}
	\caption{\label{fig:ni56_nse_mass_ratio} Estimated fraction by mass of IGE material producing $^{56}$Ni evolving in time for C-O and Hybrid C-O-Ne WD Realizations. The time-averaged value among C-O realizations is shown in yellow, with the full range of values at any point in time for the C-O realizations shown in red. Likewise, the time-averaged value among C-O-Ne realizations is shown in blue and their range of values shown in green.}
\end{figure}

\begin{figure}[!ht]
	\includegraphics[width=\linewidth]{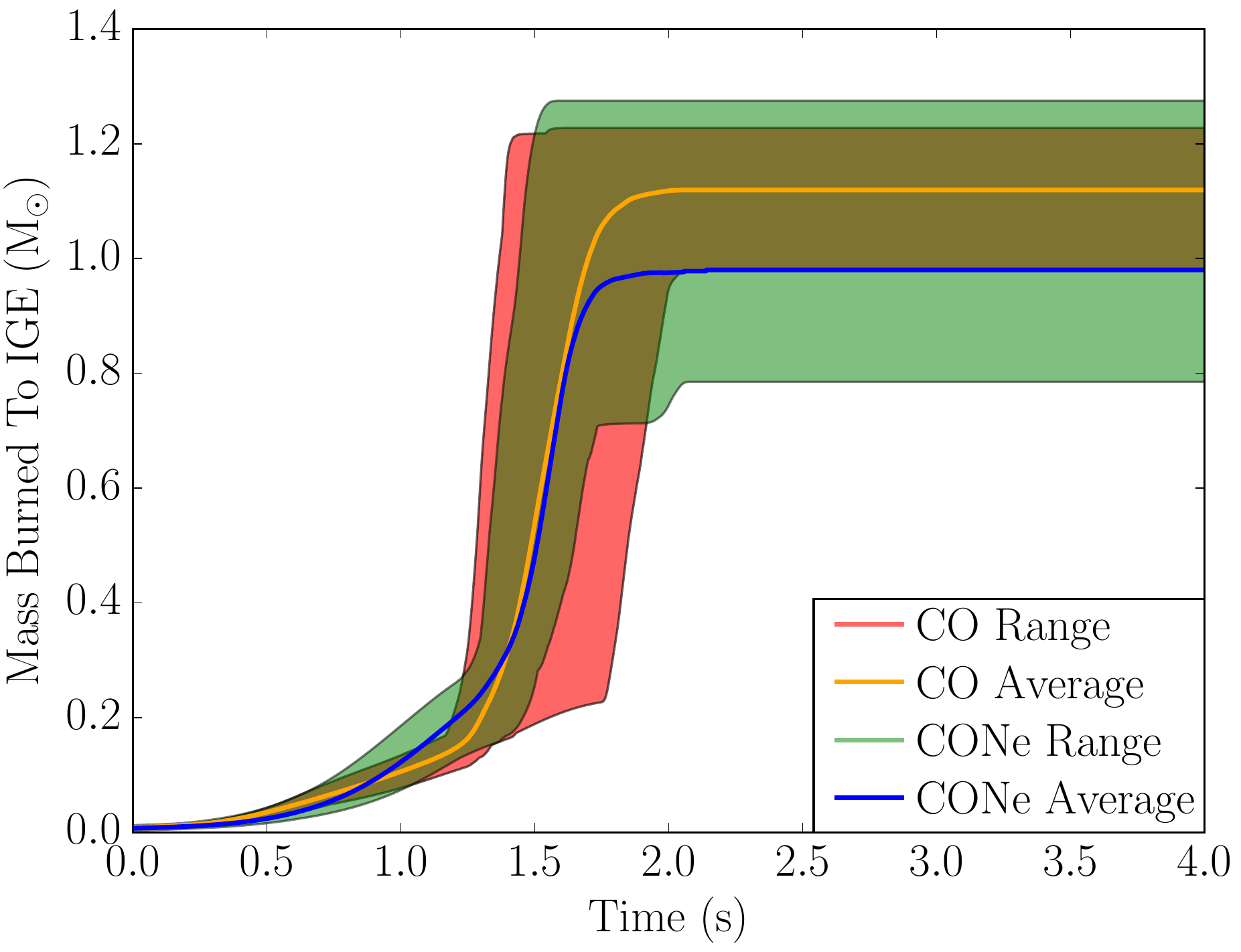}
	\caption{\label{fig:nse_burn_mass} Mass burned to IGE For C-O and Hybrid C-O-Ne WD Realizations evolving in time. The time-averaged value among C-O realizations is shown in yellow, with the full range of values at any point in time for the C-O realizations shown in red. Likewise, the time-averaged value among C-O-Ne realizations is shown in blue and their range of values shown in green.}
\end{figure}

Production of \Ni{56} is comparable between the C-O and hybrid cases
(\figref{fig:est_ni_mass}), with the full range of values from each
suite of simulations shown in the shaded regions and the mean values
shown by solid curves. The DDT event can be distinguished in the
\Ni{56} evolution by the sharp increase in the rate of \Ni{56}
production around $1.5~\second$ that rapidly yields over $0.5~\Msun$ of
\Ni{56}. While the C-O cases show a wider variation in the time at
which the DDT occurs, these also have a narrower spread in final
\Ni{56} mass relative to the hybrid models. The hybrid models also
tend to produce more \Ni{56} in the deflagration phase and some of
them show a temporary plateau in \Ni{56} production between $1.5$~s
and $2$~s. The same feature is also evident in the binding energy
curves of \figref{fig:e_binding}, computed by summing the
realization's gravitational potential, internal, and kinetic energies.

This feature is a peculiarity of the off-center ignition in the hybrid models that is 
absent in the C-O cases and results from the relatively \C{12}-poor,
cooler core region burning about $0.25$~s after the detonation front has
swept through the rest of the star. This delayed burning is shown in
Figure \ref{fig:cone_delayed_core}, which demonstrates the progression of
the detonation front into the core. Although a feature evolving over so
short a time this early in the explosion will likely not be visible in
the supernova light curves, the delayed contribution
of the core to \Ni{56} production may modify the \Ni{56} distribution in
space and velocity, potentially yielding spectral differences compared to non-delayed
hybrid as well as C-O white dwarf explosions.

The dynamical qualities of the explosion shown in the binding energy
curves of Figure \ref{fig:e_binding} indicate that the time
distribution of unbinding is more narrow for the hybrid models than
for the C-O models, though the hybrid models have a wider distribution
of final binding energies in all cases lower than the binding energies
of the C-O models within $1$~s of becoming unbound. This should
correlate to a lower expansion velocity of the ejecta, thus slower
cooling and delayed transparency relative to ejecta from C-O models.
The binding energy curves also explain the differences in expansion
the models undergo during deflagration and detonation, shown by the
mass above the density threshold $2 \times 10^7$~\grampercc\ in Figure
\ref{fig:mass_dens_lt_2e7}. For times prior to $\approx 1.2$~\second,
the C-O mass curves lie on average slightly lower than the hybrid mass
curves, indicative of a greater degree of expansion on average for the
C-O models. However, the hybrid mass curve range encompasses that of
the C-O mass curves until $\approx 1.4$~\second, reflective of the
fact that until then, some hybrid realizations
are more tightly bound than all the C-O
realizations due to burning less mass and thus expanding less. 
During the detonation phase, however, the
C-O models show a much wider variation in expansion than do the hybrid
models in spite of having a smaller range of kinetic energies and
mass burned to IGE (Figures \ref{fig:e_binding} and
\ref{fig:nse_burn_mass} below) once unbound. This is due to the C-O
models demonstrating a much wider range of DDT times than the hybrid
models.

Figure \ref{fig:ige_vs_highrho_ddt} compares the final IGE yield of
the C-O and C-O-Ne models with the degree by which the models expand
during the deflagration phase. The latter is characterized by the mass
above $2 \times 10^7$~\grampercc\ at the DDT time, with more high-density mass
indicating less expansion during deflagration. The averages of both
the C-O and C-O-Ne suites along both axes are indicated by the shaded
regions with $\pm1\sigma$ widths. The trend for both C-O and C-O-Ne
models is that less expansion during the deflagration phase results in
greater IGE yields, expected because low expansion results in there
being more high density fuel for the detonation to consume. In
addition, both the C-O and C-O-Ne models expand over similar ranges
during deflagration on average, showing they are dynamically comparable in spite of
having qualitatively different deflagration ignition
geometries. Furthermore, for similar deflagration expansion, the C-O
models tend to yield consistently greater IGE mass, suggesting that
the lower IGE yields from C-O-Ne models is not a result of these
models expanding differently than the C-O models. Rather, we interpret
this disparity as indicating that the lower IGE yield in C-O-Ne models
results from their lower \C{12} abundance and the fact that given
similar fuel density, their \Ne{20}-rich fuel will burn to cooler
temperatures than fuel in the C-O models. This in turn will result in
slower burning to IGE and thus a lower IGE yield.

The estimated \Ni{56} yields are shown in
\figref{fig:ni56_vs_nse_mass} across the range of masses burned to IGE
for all C-O and C-O-Ne realizations at $4.0$~\second\ simulation time,
at which point the total mass burned to \Ni{56} had become constant,
c.f.\ \figref{fig:est_ni_mass}. For comparable masses burned to IGE,
the hybrid models tend to consistently produce slightly more \Ni{56}
than the C-O models, although the ratio of IGE mass producing \Ni{56}
given by the slope is the same in both cases, within the fit error.
The reason for this trend is evident from Figure
\ref{fig:ni56_nse_mass_ratio}, which shows the fraction by mass of IGE
material producing \Ni{56} evolving in time, and Figure
\ref{fig:nse_burn_mass}, which shows the concurrent evolution of mass
burned to IGE. During the deflagration phase, the C-O-Ne models on
average burn more material to IGE and also had a significantly higher
fraction of IGE material producing \Ni{56}, yielding more \Ni{56} than
the C-O models. This may be due to greater neutronization in the early
deflagration of the C-O models, which are ignited closer to the center
and thus at slightly higher density than the initial deflagration of
the C-O-Ne models. However, during the subsequent detonation phase,
the C-O models on average burn more mass to IGE while maintaining a
\Ni{56}/IGE fraction very similar to that of the C-O-Ne models,
yielding significantly more \Ni{56} by the end of the detonation
phase.  For reference, \tabref{YieldSummaryTable} summarizes the
average values of \Ni{56} yield, IGE yield, and final kinetic energy
for the 35 C-O and 35 C-O-Ne realizations with one standard deviation
uncertainties.

\begin{table}[htbp]
  \caption{Average Yields and Kinetic Energy}
  \begin{center}
    \begin{tabular}{llll}
      \vspace{-0.75cm} \\ \hline \hline
      Progenitor & \Ni{56} & IGE & Kinetic Energy \\
      Type & (\Msun) & (\Msun) & $\left(\times 10^{51} ~\erg\right)$ \\ \hline
      C-O & $0.97 \pm 0.06$ & $1.12 \pm 0.07$ & $1.39 \pm 0.05$ \\
      C-O-Ne & $0.86 \pm 0.10$ & $0.98 \pm 0.11$ & $1.06 \pm 0.10$ \\ \hline
    \end{tabular}
  \end{center}
  \label{YieldSummaryTable}
\end{table}

\section{Conclusions}
\label{sec:conclusions}
Our simulations of thermonuclear (Type Ia) supernovae from both hybrid
C-O-Ne and reference C-O white dwarf progenitors using the deflagration
to detonation transition paradigm have shown that on average the
hybrid progenitors yield $0.1~\Msun$ less \Ni{56} than the C-O white
dwarfs. While this indicates that Type Ia supernovae from C-O-Ne
hybrids will be dimmer on average than those from C-O white dwarfs, we
also find sufficient variance in burning efficiency with the geometry
of the ignition region precipitating thermonuclear runaway such that
there are some hybrid progenitors that yield more \Ni{56} than some
C-O progenitors. Furthermore, we have found that not only do hybrid
C-O-Ne progenitors deposit an average of 24\% less kinetic energy in
their ejecta than C-O progenitors but also this trend of more weakly
expelled ejecta from hybrids is robust across all ignition geometries.
The consistency of this result suggests it is a consequence of the
lower energy release from Ne burning compared to C burning in spite of
the fact that using \Ne{20} as an alternate fuel can still yield
comparable \Ni{56} production in some cases.

As we noted above, we found considerable variation in the \Ni{56}
production for both hybrid and traditional C-O models, and in
particular, we found a much wider range of DDT times in the C-O models
than the hybrid models. While in the realm of speculation, this result
could follow from a greater degree of randomization in the geometry of
the initially burned region for the C-O models than in the hybrids. The
C-O models are initialized with an amplitude perturbation of the
initially burned region comprised of several angular modes, whereas the
thickness of the initially burned region in each of the hybrid models
is controlled by a single angular mode.

We also found that for some ignition geometries in the hybrid
progenitor, a combination of off-center ignition, flame buoyancy, and
composition permits their cooler core region to delay burning until
nearly 0.25~\second\ after the detonation front has consumed the rest
of the star. This result is unique to the hybrid progenitors that rely
on the Urca process to provide a lower bound on the convective zone
outside the core of the white dwarf. Delayed core burning in these
white dwarfs may result in a modified \Ni{56} distribution in their
ejecta compared to ejecta from C-O white dwarfs or even other hybrids
with prompt core burning. Exploration of such effects is the subject
of future work.

As our explosions from hybrid progenitors have a lower \Ni{56} yield
and hence lower brightness than traditional C-O models, the question
of these events as the source of observed subluminous events, e.g.
type Iax supernovae~\citep{foleyetal2013} arises. Our finding of an
average \Ni{56} yield of $0.1~\Msun$ less than the C-O, (and the larger
range of yields) indicates that explosions from these progenitors are
not subluminous and cannot on their own explain subluminous events like
type Iax supernovae.

A recent study by \citet{kromeretal2015} addressed pure deflagrations
in near-Chandrasekhar-mass hybrid WDs as the possible progenitor
systems of these faint events. The study found that most of the mass
stays bound and that early epoch light curves and spectra calculated
from the explosion models are consistent with observations of SN
2008ha~\citep{foleyetal2009}. We note that comparison between our
results and these is difficult for reasons besides the obvious
difference of the detonation phase in our simulations. The
near-Chandrasekhar-mass progenitor model of \citet{kromeretal2015} is
substantially different in that it is parameterized and it does not
include the effects of late-time convection, or the URCA process.
Also, the ignition of the deflagration is substantially different. For
these reasons, there is limited utility in a direct comparison between
results.


\acknowledgements

This work was supported in part by the Department of Energy under
grant DE-FG02-87ER40317. The software used in this work was in part
developed by the DOE-supported ASC/Alliances Center for Astrophysical
Thermonuclear Flashes at the University of Chicago. Results in this
paper were obtained using the high-performance computing system at the
Institute for Advanced Computational Science at Stony Brook
University. The authors thank Sam Jones and Ivo Seitenzahl for
fruitful discussions at the Fifty One Ergs conference, June 1-6, 2015.

\bibliography{master}

\end{document}